\pdfoutput=1

\documentclass[12pt,a4paper]{article}

\usepackage{ifthen} 
\newboolean{pdflatex}
\setboolean{pdflatex}{true} 

\newboolean{articletitles}
\setboolean{articletitles}{true} 

\newboolean{uprightparticles}
\setboolean{uprightparticles}{false} 

\newboolean{inbibliography}
\setboolean{inbibliography}{false} 

\usepackage[top=1in, bottom=1.25in, left=1in, right=1in]{geometry}

\usepackage{microtype}
\usepackage{lineno} 
\usepackage{xspace}
\usepackage{caption}

\usepackage{graphicx}  
\usepackage{color}
\usepackage{colortbl}
\graphicspath{{./figs/}} 
\usepackage{rotating}
\usepackage{subfig}
\usepackage{amsmath} 
\usepackage{amssymb}
\usepackage{amsfonts}
\usepackage{upgreek} 
\newcommand*\patchAmsMathEnvironmentForLineno[1]{%
\expandafter\let\csname old#1\expandafter\endcsname\csname #1\endcsname
\expandafter\let\csname oldend#1\expandafter\endcsname\csname
end#1\endcsname
 \renewenvironment{#1}%
   {\linenomath\csname old#1\endcsname}%
   {\csname oldend#1\endcsname\endlinenomath}%
}
\newcommand*\patchBothAmsMathEnvironmentsForLineno[1]{%
  \patchAmsMathEnvironmentForLineno{#1}%
  \patchAmsMathEnvironmentForLineno{#1*}%
}

\AtBeginDocument{%
\patchBothAmsMathEnvironmentsForLineno{equation}%
\patchBothAmsMathEnvironmentsForLineno{align}%
\patchBothAmsMathEnvironmentsForLineno{flalign}%
\patchBothAmsMathEnvironmentsForLineno{alignat}%
\patchBothAmsMathEnvironmentsForLineno{gather}%
\patchBothAmsMathEnvironmentsForLineno{multline}%
\patchBothAmsMathEnvironmentsForLineno{eqnarray}%
}

\usepackage{hyperref}    
\usepackage[all]{hypcap} 

\usepackage{xspace} 
\usepackage{upgreek}

\def\lhcb {\mbox{LHCb}\xspace}

\def\babar  {\mbox{BaBar}\xspace}

\def\velo   {VELO\xspace}
\def\rich   {RICH\xspace}

\def\MagUp {\mbox{\em Mag\kern -0.05em Up}\xspace}

\ifthenelse{\boolean{uprightparticles}}%
{
 
 \def\Pgamma      {\ensuremath{\upgamma}\xspace}

 \def\Ppi         {\ensuremath{\uppi}\xspace}

 \def\PDelta      {\ensuremath{\Delta}\xspace}                 
 \def\PXi      {\ensuremath{\Xi}\xspace}                 
 \def\PLambda      {\ensuremath{\Lambda}\xspace}                 
 \def\PSigma      {\ensuremath{\Sigma}\xspace}                 
 \def\POmega      {\ensuremath{\Omega}\xspace}                 
 \def\PUpsilon      {\ensuremath{\Upsilon}\xspace}                 
 
 \def\PB      {\ensuremath{\mathrm{B}}\xspace}                 
                  
 \def\PD      {\ensuremath{\mathrm{D}}\xspace}

 \def\PK      {\ensuremath{\mathrm{K}}\xspace}

 \def\Pb      {\ensuremath{\mathrm{b}}\xspace}                 
 \def\Pc      {\ensuremath{\mathrm{c}}\xspace}                 
 \def\Pd      {\ensuremath{\mathrm{d}}\xspace}

 \def\Pi      {\ensuremath{\mathrm{i}}\xspace}

 \def\Ps      {\ensuremath{\mathrm{s}}\xspace}                 
 \def\Pt      {\ensuremath{\mathrm{t}}\xspace}                 
 \def\Pu      {\ensuremath{\mathrm{u}}\xspace}

}
{
 
 \def\Pgamma      {\ensuremath{\gamma}\xspace}

 \def\Ppi         {\ensuremath{\pi}\xspace}

 \mathchardef\PDelta="7101
 \mathchardef\PXi="7104
 \mathchardef\PLambda="7103
 \mathchardef\PSigma="7106
 \mathchardef\POmega="710A
 \mathchardef\PUpsilon="7107
                  
 \def\PB      {\ensuremath{B}\xspace}                 
                  
 \def\PD      {\ensuremath{D}\xspace}

 \def\PK      {\ensuremath{K}\xspace}

 \def\Pb      {\ensuremath{b}\xspace}                 
 \def\Pc      {\ensuremath{c}\xspace}                 
 \def\Pd      {\ensuremath{d}\xspace}

 \def\Pi      {\ensuremath{i}\xspace}

 \def\Ps      {\ensuremath{s}\xspace}                 
 \def\Pt      {\ensuremath{t}\xspace}                 
 \def\Pu      {\ensuremath{u}\xspace}

}

\makeatletter
\ifcase \@ptsize \relax
  \newcommand{\miniscule}{\@setfontsize\miniscule{4}{5}}
\or
  \newcommand{\miniscule}{\@setfontsize\miniscule{5}{6}}
\or
  \newcommand{\miniscule}{\@setfontsize\miniscule{5}{6}}
\fi
\makeatother

\DeclareRobustCommand{\optbar}[1]{\shortstack{{\miniscule (\rule[.5ex]{1.25em}{.18mm})}
  \\ [-.7ex] $#1$}}

\def\uquark    {{\ensuremath{\Pu}}\xspace}
\def\uquarkbar {{\ensuremath{\overline \uquark}}\xspace}

\def\dquark    {{\ensuremath{\Pd}}\xspace}

\def\squark    {{\ensuremath{\Ps}}\xspace}

\def\cquark    {{\ensuremath{\Pc}}\xspace}
\def\cquarkbar {{\ensuremath{\overline \cquark}}\xspace}

\def\bquark    {{\ensuremath{\Pb}}\xspace}

\def\tquark    {{\ensuremath{\Pt}}\xspace}

\def\pion   {{\ensuremath{\Ppi}}\xspace}
\def\piz    {{\ensuremath{\pion^0}}\xspace}

\def\pip    {{\ensuremath{\pion^+}}\xspace}
\def\pim    {{\ensuremath{\pion^-}}\xspace}
\def\pipm   {{\ensuremath{\pion^\pm}}\xspace}
\def\pimp   {{\ensuremath{\pion^\mp}}\xspace}

\def\kaon    {{\ensuremath{\PK}}\xspace}
  \def\Kbar    {{\kern 0.2em\overline{\kern -0.2em \PK}{}}\xspace}

\def\KorKbar    {\kern 0.18em\optbar{\kern -0.18em K}{}\xspace}

\def\Kp      {{\ensuremath{\kaon^+}}\xspace}
\def\Km      {{\ensuremath{\kaon^-}}\xspace}
\def\Kpm     {{\ensuremath{\kaon^\pm}}\xspace}
\def\Kmp     {{\ensuremath{\kaon^\mp}}\xspace}
\def\KS      {{\ensuremath{\kaon^0_{\mathrm{ \scriptscriptstyle S}}}}\xspace}

\def\Kstarz  {{\ensuremath{\kaon^{*0}}}\xspace}

\def\Kstar   {{\ensuremath{\kaon^*}}\xspace}

\def\Kstarp  {{\ensuremath{\kaon^{*+}}}\xspace}
\def\Kstarm  {{\ensuremath{\kaon^{*-}}}\xspace}
\def\Kstarpm {{\ensuremath{\kaon^{*\pm}}}\xspace}

  \def\Dbar    {{\kern 0.2em\overline{\kern -0.2em \PD}{}}\xspace}
\def\D       {{\ensuremath{\PD}}\xspace}

\def\DorDbar    {\kern 0.18em\optbar{\kern -0.18em D}{}\xspace}
\def\Dz      {{\ensuremath{\D^0}}\xspace}
\def\Dzb     {{\ensuremath{\Dbar{}^0}}\xspace}

\def\Dstar   {{\ensuremath{\D^*}}\xspace}

\def\Dstarz  {{\ensuremath{\D^{*0}}}\xspace}

\def\Dstarp  {{\ensuremath{\D^{*+}}}\xspace}

\def\B       {{\ensuremath{\PB}}\xspace}
\def\Bbar    {{\ensuremath{\kern 0.18em\overline{\kern -0.18em \PB}{}}}\xspace}

\def\BorBbar    {\kern 0.18em\optbar{\kern -0.18em B}{}\xspace}
\def\Bz      {{\ensuremath{\B^0}}\xspace}

\def\Bu      {{\ensuremath{\B^+}}\xspace}
\def\Bub     {{\ensuremath{\B^-}}\xspace}
\def\Bp      {{\ensuremath{\Bu}}\xspace}
\def\Bm      {{\ensuremath{\Bub}}\xspace}
\def\Bpm     {{\ensuremath{\B^\pm}}\xspace}

\def\Bs      {{\ensuremath{\B^0_\squark}}\xspace}

\def\Bdb     {{\ensuremath{\Bbar{}^0}}\xspace}

  \def\Y#1S{\ensuremath{\PUpsilon{(#1S)}}\xspace}

\def\Lz          {{\ensuremath{\PLambda}}\xspace}
\def\Lbar        {{\ensuremath{\kern 0.1em\overline{\kern -0.1em\PLambda}}}\xspace}
\def\LorLbar    {\kern 0.18em\optbar{\kern -0.18em \PLambda}{}\xspace}

\def\Lb      {{\ensuremath{\Lz^0_\bquark}}\xspace}

\def\Lc      {{\ensuremath{\Lz^+_\cquark}}\xspace}

\def\BF         {{\ensuremath{\mathcal{B}}}\xspace}

\def\BR         {\BF}
\newcommand{\decay}[2]{\ensuremath{#1\!\to #2}\xspace}         

\def\to                 {\ensuremath{\rightarrow}\xspace}

\def\order   {{\ensuremath{\mathcal{O}}}\xspace}

\def\CP                {{\ensuremath{C\!P}}\xspace}

\def\Vud  {{\ensuremath{V_{\uquark\dquark}}}\xspace}
\def\Vcd  {{\ensuremath{V_{\cquark\dquark}}}\xspace}
\def\Vtd  {{\ensuremath{V_{\tquark\dquark}}}\xspace}

\def\Vub  {{\ensuremath{V_{\uquark\bquark}}}\xspace}
\def\Vcb  {{\ensuremath{V_{\cquark\bquark}}}\xspace}

\def\Vubs  {{\ensuremath{V_{\uquark\bquark}^\ast}}\xspace}
\def\Vcbs  {{\ensuremath{V_{\cquark\bquark}^\ast}}\xspace}
\def\Vtbs  {{\ensuremath{V_{\tquark\bquark}^\ast}}\xspace}

\def\AT#1     {\ensuremath{A_{\mathrm{T}}^{#1}}\xspace}           

\def\C#1      {\ensuremath{\mathcal{C}_{#1}}\xspace}                       
\def\Cp#1     {\ensuremath{\mathcal{C}_{#1}^{'}}\xspace}                    
\def\Ceff#1   {\ensuremath{\mathcal{C}_{#1}^{\mathrm{(eff)}}}\xspace}        
\def\Cpeff#1  {\ensuremath{\mathcal{C}_{#1}^{'\mathrm{(eff)}}}\xspace}       
\def\Ope#1    {\ensuremath{\mathcal{O}_{#1}}\xspace}                       
\def\Opep#1   {\ensuremath{\mathcal{O}_{#1}^{'}}\xspace}                    

\newcommand{\tev}{\ifthenelse{\boolean{inbibliography}}{\ensuremath{~T\kern -0.05em eV}\xspace}{\ensuremath{\mathrm{\,Te\kern -0.1em V}}}\xspace}
\newcommand{\gev}{\ensuremath{\mathrm{\,Ge\kern -0.1em V}}\xspace}
\newcommand{\mev}{\ensuremath{\mathrm{\,Me\kern -0.1em V}}\xspace}
\newcommand{\kev}{\ensuremath{\mathrm{\,ke\kern -0.1em V}}\xspace}
\newcommand{\ev}{\ensuremath{\mathrm{\,e\kern -0.1em V}}\xspace}
\newcommand{\gevc}{\ensuremath{{\mathrm{\,Ge\kern -0.1em V\!/}c}}\xspace}
\newcommand{\mevc}{\ensuremath{{\mathrm{\,Me\kern -0.1em V\!/}c}}\xspace}
\newcommand{\gevcc}{\ensuremath{{\mathrm{\,Ge\kern -0.1em V\!/}c^2}}\xspace}
\newcommand{\gevgevcccc}{\ensuremath{{\mathrm{\,Ge\kern -0.1em V^2\!/}c^4}}\xspace}
\newcommand{\mevcc}{\ensuremath{{\mathrm{\,Me\kern -0.1em V\!/}c^2}}\xspace}

\def\mum  {\ensuremath{{\,\upmu\mathrm{m}}}\xspace}

\def\invfb   {\ensuremath{\mbox{\,fb}^{-1}}\xspace}

\def\order{{\ensuremath{\mathcal{O}}}\xspace}
\newcommand{\chisq}{\ensuremath{\chi^2}\xspace}

\newcommand{\chisqip}{\ensuremath{\chi^2_{\text{IP}}}\xspace}

\def\gsim{{~\raise.15em\hbox{$>$}\kern-.85em
          \lower.35em\hbox{$\sim$}~}\xspace}
\def\lsim{{~\raise.15em\hbox{$<$}\kern-.85em
          \lower.35em\hbox{$\sim$}~}\xspace}

\def\ptot       {\mbox{$p$}\xspace}
\def\pt         {\mbox{$p_{\mathrm{ T}}$}\xspace}

\def\evtgen     {\mbox{\textsc{EvtGen}}\xspace}

\def\geant      {\mbox{\textsc{Geant4}}\xspace}

\def\photos     {\mbox{\textsc{Photos}}\xspace}

\def\pythia     {\mbox{\textsc{Pythia}}\xspace}

\def\tell1  {TELL1\xspace}
\def\ukl1   {UKL1\xspace}

\newcommand{\eg}{\mbox{\itshape e.g.}\xspace}

\usepackage{cite} 
\usepackage{mciteplus}
\usepackage{longtable} 

\begin{document}

\renewcommand{\thefootnote}{\fnsymbol{footnote}}
\setcounter{footnote}{1}


\begin{titlepage}
\pagenumbering{roman}

\vspace*{-1.5cm}
\centerline{\large EUROPEAN ORGANIZATION FOR NUCLEAR RESEARCH (CERN)}
\vspace*{1.5cm}
\noindent
\begin{tabular*}{\linewidth}{lc@{\extracolsep{\fill}}r@{\extracolsep{0pt}}}
\ifthenelse{\boolean{pdflatex}}
{\vspace*{-2.7cm}\mbox{\!\!\!\includegraphics[width=.14\textwidth]{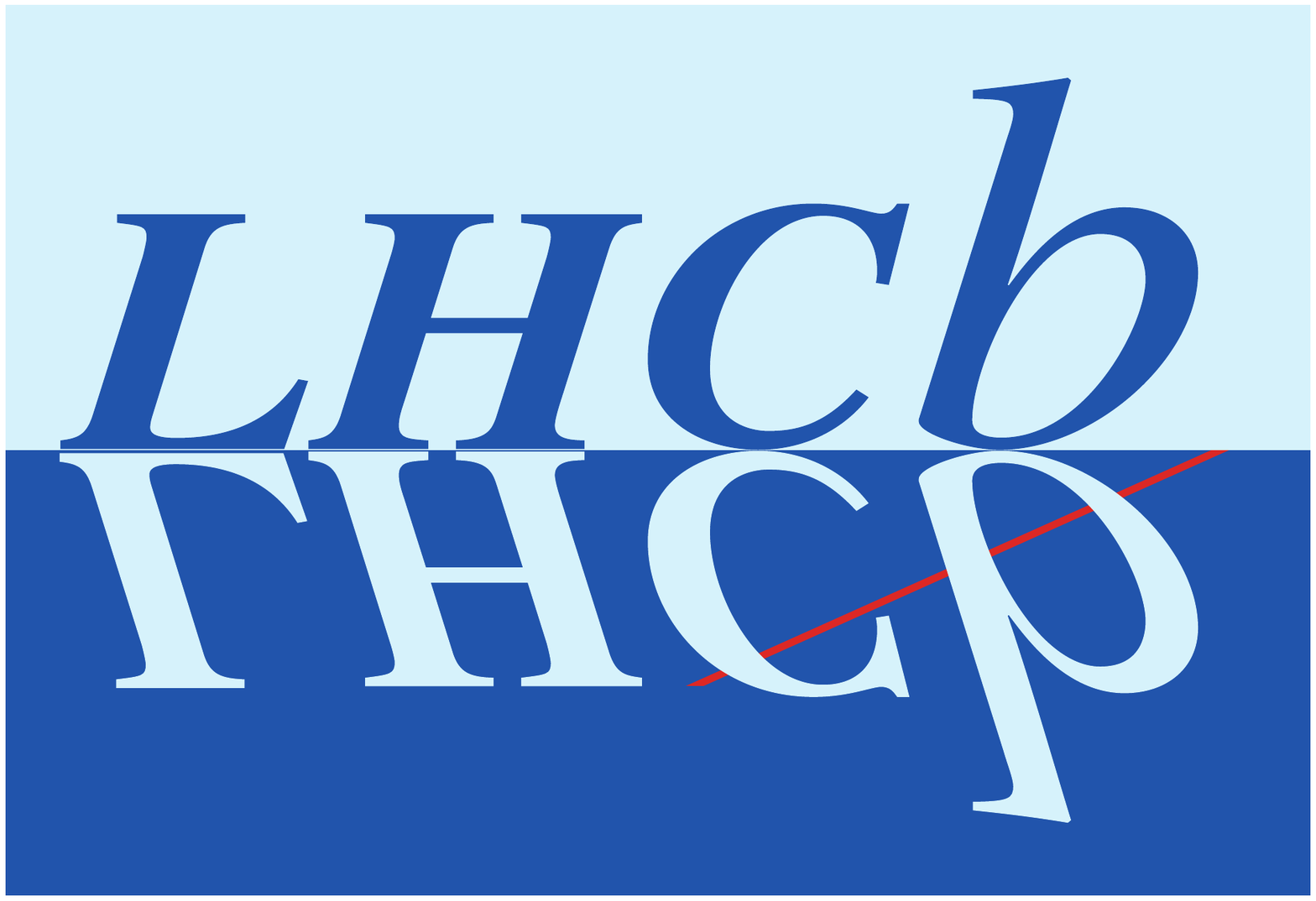}} & &}%
{\vspace*{-1.2cm}\mbox{\!\!\!\includegraphics[width=.12\textwidth]{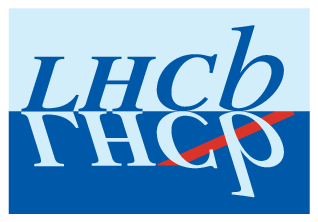}} & &}%
\\
 & & CERN-EP-2017-208 \\  
 & & LHCb-PAPER-2017-030 \\  
 & & 26th March 2018 \\ 
\end{tabular*}

\vspace*{4.0cm}

{\normalfont\bfseries\boldmath\huge
\begin{center}
  Measurement of \CP observables in \decay{\Bpm}{\D\Kstarpm} decays using two- and four-body \D final states
\end{center}
}

\vspace*{2.0cm}

\begin{center}
The LHCb collaboration\footnote{Authors are listed at the end of this paper.}
\end{center}

\vspace{\fill}

\begin{abstract}
\noindent 
Measurements of \CP observables in \decay{\Bpm}{\D\Kstarpm} decays are presented, where \D denotes a superposition of \Dz and \Dzb meson states. Decays of the \D meson to \Km\pip, \Km\Kp, \pim\pip, \Km\pip\pim\pip and \pim\pip\pim\pip are used and the \Kstarpm meson is reconstructed in the \KS\pipm final state. This analysis uses a data sample of $pp$ collisions collected with the \lhcb experiment, corresponding to integrated luminosities of 1\invfb, 2\invfb and 1.8\invfb at centre-of-mass energies $\sqrt{s} = $ 7\tev, 8\tev and 13\tev, respectively. The sensitivity of the results to the CKM angle \Pgamma is discussed.

\end{abstract}

\vspace*{2.0cm}

\begin{center}
  Published in JHEP 11 (2017) 156
\end{center}

\vspace{\fill}

{\footnotesize 
\centerline{\copyright~CERN on behalf of the \lhcb collaboration, licence \href{http://creativecommons.org/licenses/by/4.0/}{CC-BY-4.0}.}}
\vspace*{2mm}

\end{titlepage}


\newpage
\setcounter{page}{2}
\mbox{~}

\cleardoublepage

\renewcommand{\thefootnote}{\arabic{footnote}}
\setcounter{footnote}{0}

\cleardoublepage


\pagestyle{plain} 
\setcounter{page}{1}
\pagenumbering{arabic}


\section{Introduction}
\label{sec:Introduction}

A key characteristic of the Standard Model is that \CP violation originates from a single phase in the CKM quark-mixing matrix~\cite{Cabibbo,KM}. In the Standard Model the CKM matrix is unitary, leading to the condition $\Vud\Vubs + \Vcd\Vcbs + \Vtd\Vtbs = 0$, where $V_{ij}$ are the CKM matrix elements. This relation is represented as a triangle in the complex plane, with angles $\alpha$, $\beta$ and \Pgamma, and an area proportional to the amount of \CP violation in the quark sector of the Standard Model~\cite{CKMtriangle}. Overconstraining this unitarity triangle may lead to signs of physics beyond the Standard Model. The CKM angle $\Pgamma \equiv \arg\left(-\frac{\Vud{\Vub}^*}{\Vcd{\Vcb}^*}\right)$ is the least well-known angle of the CKM unitarity triangle. The latest published \lhcb combination from direct measurements with charged and neutral \B decays to a \D meson (reconstructed in one of a variety of final states) and a kaon is $\Pgamma = \left(72.2^{+6.8}_{-7.3}\right)^{\circ}$~\cite{LHCB-PAPER-2016-032}. A global fit to the CKM triangle by the CKMfitter group~\cite{CKMfitter} obtains a \Pgamma value of $(66.9^{+0.9}_{-3.4})^{\circ}$, where this determination of \Pgamma excludes all direct measurements. The uncertainties on the indirect measurement are expected to decrease as lattice QCD calculations become more accurate. Therefore, precision at the level of $1^\circ$ on a direct measurement of \Pgamma would test the consistency of the direct and indirect measurements and thereby the Standard Model. This precision can be achieved through a combination of measurements of various \B decays that are sensitive to \Pgamma.

Direct measurements of \Pgamma can be made by exploiting the interference between \decay{\bquark}{\cquark\uquarkbar\squark} and \decay{\bquark}{\uquark\cquarkbar\squark} transitions. These transitions are present in $\B \to \D^{(*)}\kaon^{(*)}$ decays. This analysis measures \CP violation in $\B^- \to \D \Kstar(892)^-$ decays,\footnote{The inclusion of charge-conjugate processes is implied, except when discussing ratios or asymmetries between \Bp and \Bm decays.} with  \mbox{\decay{\Kstar(892)^-}{\KS(\pip\pim)\pim}}, where \D denotes a superposition of \Dz and \Dzb meson states. In this paper \Kstarm is used to represent the $\Kstar(892)^-$ resonance. The effect of the interference is observed by reconstructing the \D meson in a final state accessible to both \Dz and \Dzb meson states, which gives sensitivity to the weak phase \Pgamma. In this analysis, only \D mesons decaying to two or four charged kaons and/or pions are considered. The branching fraction of \decay{\Bm}{\D\Kstarm} is of a similar magnitude to \decay{\Bm}{\D\Km}, which has been extensively analysed at \lhcb~\cite{LHCB-PAPER-2016-003, LHCB-PAPER-2014-041, LHCB-PAPER-2015-014}. However, the reconstruction efficiencies associated with the \decay{\Kstarm}{\KS\pim} decay are lower due to the presence of a long-lived neutral particle.

Two main classes of \D decays are used. The first employs \D decays into the \CP-even eigenstates \Kp\Km and \pip\pim; these are referred to here as the ``GLW'' decay modes~\cite{GL,GW}. The second class of decay modes involves \D decays to \Kmp\pipm, which is not a \CP eigenstate. In the favoured decay, the pion from the \D meson and that from the \Kstarm meson have opposite charge, while in the suppressed decay (referred to here as the ``ADS'' ~\cite{ADS,ADS-2001} decay mode) the pion from the \D meson that from the \Kstarm meson have the same charge. The favoured mode is used  as a control mode for many aspects of the analysis since no \CP asymmetry is expected. The ADS decay mode is a combination of a CKM-favoured \decay{\Bm}{\Dz\Kstarm} decay, followed by a doubly Cabibbo-suppressed \decay{\Dz}{\Kp\pim} decay, and a CKM- and colour-suppressed \decay{\Bm}{\Dzb\Kstarm} decay, followed by a Cabibbo-favoured \decay{\Dzb}{\Kp\pim} decay. Both paths to the same final state have amplitudes of similar size, and interference effects are therefore magnified in comparison to the GLW decay modes, where the decay path via the CKM-favoured \decay{\Bm}{\Dz\Kstarm} dominates. Studies of \decay{\Bm}{D\Km} and \decay{\Bz}{\D\Kstarz} decays have been published by the \lhcb collaboration~\cite{LHCB-PAPER-2016-003, LHCB-PAPER-2014-028}. 

The GLW and ADS methods can be extended to the \decay{\D}{\Kmp\pipm\pimp\pipm} and \mbox{\decay{\D}{\pip\pim\pip\pim}} inclusive four-body final states, provided external information is available on the overall behaviour of the intermediate resonances, averaged over phase space~\cite{charminfo,charm4pi}. These channels have previously been studied for \decay{\Bm}{\D\Km} decays~\cite{LHCB-PAPER-2016-003}, and are included in this paper for the first time in \decay{\Bm}{\D\Kstarm} decays. The \decay{\Bm}{\D\Kstarm} channel has previously been investigated by the BaBar collaboration using a variety of two-body \D decay modes~\cite{BaBarDKstar}. 
Also, both the BaBar and Belle collaborations have performed studies on \decay{\Bm}{\D\Kstarm} with \decay{\D}{\KS\pip\pim}~\cite{BaBarGGSZ,BelleGGSZ}.

Twelve quantities, collectively referred to as \CP observables, are measured in this analysis

\begin{itemize}
\item{The \CP asymmetry for the favoured decay mode
\begin{equation}
A_{\kaon\pi} = \frac{\Gamma\left(\decay{\Bm}{\D(\Km\pip)\Kstarm}\right) - \Gamma\left(\decay{\Bp}{\D(\Kp\pim)\Kstarp}\right)}{\Gamma\left(\decay{\Bm}{\D(\Km\pip)\Kstarm}\right) + \Gamma\left(\decay{\Bp}{\D(\Kp\pim)\Kstarp}\right)} \text{ .}
\label{eqn:Akpi}
\end{equation}}
\item{The \CP asymmetry for the \decay{\D}{\Kp\Km} decay mode
\begin{equation}
A_{\kaon\kaon} = \frac{\Gamma\left(\decay{\Bm}{\D(\Kp\Km)\Kstarm}\right) - \Gamma\left(\decay{\Bp}{\D(\Kp\Km)\Kstarp}\right)}{\Gamma\left(\decay{\Bm}{\D(\Kp\Km)\Kstarm}\right) + \Gamma\left(\decay{\Bp}{\D(\Kp\Km)\Kstarp}\right)} \text{ . }
\label{eqn:Akk}
\end{equation}
}
\item{The \CP asymmetry for the \decay{\D}{\pip\pim} decay mode
\begin{equation}
A_{\pi\pi} = \frac{\Gamma\left(\decay{\Bm}{\D(\pip\pim)\Kstarm}\right) - \Gamma\left(\decay{\Bp}{\D(\pip\pim)\Kstarp}\right)}{\Gamma\left(\decay{\Bm}{\D(\pip\pim)\Kstarm}\right) + \Gamma\left(\decay{\Bp}{\D(\pip\pim)\Kstarp}\right)} \text{ . }
\label{eqn:Apipi}
\end{equation}}
\item{The ratio of the rate for the \decay{\D}{\Kp\Km} decay mode to that of the favoured decay mode, scaled by the branching fractions
{\footnotesize
\begin{equation}
R_{\kaon\kaon} = \frac{\Gamma\left(\decay{\Bm}{\D(\Kp\Km)\Kstarm}\right) + \Gamma\left(\decay{\Bp}{\D(\Kp\Km)\Kstarp}\right)}{\Gamma\left(\decay{\Bm}{\D(\Km\pip)\Kstarm}\right) + \Gamma\left(\decay{\Bp}{\D(\Kp\pim)\Kstarp}\right)} \times \frac{\BR(D^0 \to K^-\pi^+)}{\BR(D^0 \to K^+K^-)} \text{ . }
\label{eqn:Rkk}
\end{equation}
}}
\item{The ratio of the rate for the \decay{\D}{\pip\pim} decay mode to that of the favoured decay mode, scaled by the branching fractions
{\footnotesize
\begin{equation}
R_{\pi\pi} = \frac{\Gamma\left(\decay{\Bm}{\D(\pip\pim)\Kstarm}\right) + \Gamma\left(\decay{\Bp}{\D(\pip\pim)\Kstarp}\right)}{\Gamma\left(\decay{\Bm}{\D(\Km\pip)\Kstarm}\right) + \Gamma\left(\decay{\Bp}{\D(\Kp\pim)\Kstarp}\right)} \times \frac{\BR(D^0 \to K^-\pi^+)}{\BR(D^0 \to \pi^+\pi^-)} \text{ . }
\label{eqn:Rpipi}
\end{equation}}}
\item{The ratio of the rate for the ADS decay mode to that of the favoured decay mode for \Bp decays
\begin{equation}
R^+_{K\pi} = \frac{\Gamma\left(\decay{\Bp}{\D(\Km\pip)\Kstarp}\right)}{\Gamma\left(\decay{\Bp}{\D(\Kp\pim)\Kstarp}\right)} \text{ . }
\label{eqn:Rplus}
\end{equation}
}
\item{The ratio of the rate for the ADS decay mode to that of the favoured decay mode for \Bm decays
\begin{equation}
R^-_{K\pi} = \frac{\Gamma\left(\decay{\Bm}{\D(\Kp\pim)\Kstarm}\right)}{\Gamma\left(\decay{\Bm}{\D(\Km\pip)\Kstarm}\right)} \text{ . }
\label{eqn:Rminus}
\end{equation}
}
\item{The \CP asymmetry for the favoured \decay{\Dz}{\Km\pip\pim\pip} decay mode
\begin{equation}
A_{\kaon\pi\pi\pi} = \frac{\Gamma\left(\decay{\Bm}{\D(\Km\pip\pim\pip)\Kstarm}\right) - \Gamma\left(\decay{\Bp}{\D(\Kp\pim\pip\pim)\Kstarp}\right)}{\Gamma\left(\decay{\Bm}{\D(\Km\pip\pim\pip)\Kstarm}\right) + \Gamma\left(\decay{\Bp}{\D(\Kp\pim\pip\pim)\Kstarp}\right)} \text{ . }
\label{eqn:Akpipipi}
\end{equation}}
\item{The \CP asymmetry for the \decay{\D}{\pip\pim\pip\pim} decay mode
\begin{equation}
A_{\pi\pi\pi\pi} = \frac{\Gamma\left(\decay{\Bm}{\D(\pip\pim\pip\pim)\Kstarm}\right) - \Gamma\left(\decay{\Bp}{\D(\pip\pim\pip\pim)\Kstarp}\right)}{\Gamma\left(\decay{\Bm}{\D(\pip\pim\pip\pim)\Kstarm}\right) + \Gamma\left(\decay{\Bp}{\D(\pip\pim\pip\pim)\Kstarp}\right)} \text{ . }
\label{eqn:Apipipipi}
\end{equation}
}
\item{The ratio of the rate for the \decay{\D}{\pip\pim\pip\pim} decay mode to that of the favoured decay mode, scaled by the branching fractions
\begin{multline}
R_{\pi\pi\pi\pi} = \frac{\Gamma\left(\decay{\Bm}{\D(\pip\pim\pip\pim)\Kstarm}\right) + \Gamma\left(\decay{\Bp}{\D(\pip\pim\pip\pim)\Kstarp}\right)}{\Gamma\left(\decay{\Bm}{\D(\Km\pip\pim\pip)\Kstarm}\right) + \Gamma\left(\decay{\Bp}{\D(\Kp\pim\pip\pim)\Kstarp}\right)} \\
\times \frac{\mathcal{B}(D^0 \to \Km\pip\pim\pip)}{\mathcal{B}(D^0 \to \pip\pim\pip\pim)} \text{ . }
\label{eqn:Rpipipipi}
\end{multline}}
\item{The ratio of the rate for the four-body ADS decay mode to that of the four-body favoured decay mode for \Bp decays
\begin{equation}
R^+_{K\pi\pi\pi} = \frac{\Gamma\left(\decay{\Bp}{\D(\Km\pip\pim\pip)\Kstarp}\right)}{\Gamma\left(\decay{\Bp}{\D(\Kp\pim\pip\pim)\Kstarp}\right)} \text{ . }
\label{eqn:Rplus4body}
\end{equation}
}
\item{The ratio of the rate of the four-body ADS decay mode to that of the four-body favoured decay mode for \Bm decays
\begin{equation}
R^-_{K\pi\pi\pi} = \frac{\Gamma\left(\decay{\Bm}{\D(\Kp\pim\pip\pim)\Kstarm}\right)}{\Gamma\left(\decay{\Bm}{\D(\Km\pip\pim\pip)\Kstarm}\right)} \text{ . }
\label{eqn:Rminus4body}
\end{equation}
}
\end{itemize}

\noindent
The asymmetries $A_{\kaon\pi}$ and $A_{\kaon\pi\pi\pi}$ should be essentially zero due to the very small interference expected in the configuration of \B and \D decays. Due to negligible direct \CP violation in \D decays~\cite{charmcpv}, the observables $A_{\kaon\kaon}$ and $A_{\pi\pi}$ should be equal and are often labelled together as $A_{\CP+}$; similarly the observables $R_{\kaon\kaon}$ and $R_{\pi\pi}$ should be equal and are labelled $R_{\CP+}$. The analogous observables to $R_{\CP+}$ and $A_{\CP+}$ for the ADS mode are $R_{ADS}$ and $A_{ADS}$. However, $R_{ADS}$ and $A_{ADS}$ are not used for the ADS decay mode, instead the ratios are measured separately for the positive and negative charges. The reason for this choice is that the uncertainty in $A_{ADS}$ depends on the value of $R_{ADS}$, therefore these observables are statistically dependent, raising problems for the low yields expected in the ADS mode. Hence the statistically independent observables $R^+_{K\pi}$ and $R^-_{K\pi}$ are preferred.


The \CP observables measured in this analysis can be related to the physics parameters to be determined, namely \Pgamma, $r_B$ and $\delta_B$. The parameter $r_B$ is the ratio of the magnitudes between the suppressed and favoured amplitudes of the \B decay and $\delta_B$ is the strong-phase difference between these amplitudes. The expected value is $r_B$ $\sim 0.1$, similar to that in the \decay{\Bm}{\D\Km} decay. Both $r_B$ and $\delta_B$ are averaged over the region of \D\KS\pim phase space corresponding to the \Kstarm selection window. A coherence factor, $\kappa$, accounts for the contribution of $\Bm \to D \KS \pim$ decays that are not due to an intermediate $K^*(892)^{-}$ resonance~\cite{Gronau2003198}, where $\kappa = 1$ denotes a pure $K^*(892)^{-}$ contribution. Given there is a negligible effect from both charm mixing~\cite{charmmixing} and \CP violation in \D decays~\cite{charmcpv}, the relationships between the \CP observables and physics parameters are given in the following equations,

\begin{equation}
A_{\CP+} = \frac{2 \kappa r_B\sin\delta_B\sin\gamma}{1 + r_B^2 + 2 \kappa r_B\cos\delta_B\cos\gamma} \text{ ,}
\label{exp_Acp}
\end{equation}

\begin{equation}
R_{\CP+} = 1 + r_B^2 + 2 \kappa r_B\cos\delta_B\cos\gamma \text{ ,}
\label{exp_Rcp}
\end{equation}

\begin{equation}
R^{\pm}_{K\pi} = \frac{r_B^2 + \left(r_D^{K\pi}\right)^2 + 2\kappa r_B r_D^{K\pi} \cos(\delta_B + \delta_D^{K\pi} \pm \gamma)}{1 + r_B^2\left(r_D^{K\pi}\right)^2 + 2\kappa r_B r_D^{K\pi} \cos(\delta_B - \delta_D^{K\pi} \pm \gamma)} \text{ ,}
\label{exp_Rpm}
\end{equation}

\begin{equation}
A_{\pi\pi\pi\pi} = \frac{2 \kappa\left(2F_{4\pi} - 1\right) r_B\sin\delta_B\sin\gamma}{1 + r_B^2 + 2 \kappa\left(2F_{4\pi} - 1\right) r_B\cos\delta_B\cos\gamma} \text{ ,}
\label{exp_A4pi}
\end{equation}

\begin{equation}
R_{\pi\pi\pi\pi} = 1 + r_B^2 + 2 \kappa\left(2F_{4\pi} - 1\right) r_B\cos\delta_B\cos\gamma \text{ ,}
\label{exp_R4pi}
\end{equation}

\begin{equation}
R^{\pm}_{K\pi\pi\pi} = \frac{r_B^2 + \left(r_D^{K3\pi}\right)^2 + 2\kappa r_B \kappa_{K3\pi} r_D^{K3\pi} \cos(\delta_B + \delta_D^{K3\pi} \pm \gamma)}{1 + \left(r_Br_D^{K3\pi}\right)^2 + 2\kappa r_B \kappa_{K3\pi} r_D^{K3\pi} \cos(\delta_B - \delta_D^{K3\pi} \pm \gamma)} \text{ .}
\label{exp_Rpm4body}
\end{equation}
These relationships depend on several parameters describing the \D decays, which are taken from existing measurements. The parameters $r_D^{K\pi}$ and $\delta_D^{K\pi}$ are the magnitude of the amplitude ratio and the strong-phase difference between the suppressed and favoured amplitudes of the \D decay, namely \decay{\Dz}{\Kp\pim} and \decay{\Dz}{\Km\pip} respectively~\cite{HFLAV16}. Similarly, the parameters $r_D^{K3\pi}$ and $\delta_D^{K3\pi}$ are the equivalent quantities for the decays \decay{\Dz}{\Kp\pim\pip\pim} and \decay{\Dz}{\Km\pip\pim\pip}, averaged over phase space~\cite{charmk3pi,LHCB-PAPER-2015-057}. Two-body \decay{\D}{\Kmp\pipm} decays are characterised by a single strong phase, however for multibody \decay{\D}{\Kmp\pipm\pimp\pipm} decays the strong phase varies over the phase space. By averaging the strong phase variation the interference effects are diluted. This effect is accounted for by the parameter $\kappa_{K3\pi}$~\cite{charmk3pi,LHCB-PAPER-2015-057}. The parameter $F_{4\pi} \sim 0.75$~\cite{charm4pi} accounts for the fact that \decay{\D}{\pip\pim\pip\pim}, though predominantly \CP even, is not a pure \CP eigenstate.

\section{Detector, online selection and simulation}
\label{sec:detector}

The \lhcb detector~\cite{Alves:2008zz,LHCb-DP-2014-002} is a single-arm forward spectrometer covering the \mbox{pseudorapidity} range $2<\eta <5$, designed for the study of particles containing \bquark or \cquark quarks. The detector includes a high-precision tracking system consisting of a silicon-strip vertex detector (\velo) surrounding the $pp$ interaction region, a large-area silicon-strip detector (TT) located upstream of a dipole magnet with a bending power of about $4{\mathrm{\,Tm}}$, and three stations of silicon-strip detectors and straw drift tubes placed downstream of the magnet. The tracking system provides a measurement of momentum, \ptot, of charged particles with a relative uncertainty that varies from 0.5\% at low momentum to 1.0\% at 200\gevc. The minimum distance of a track to a primary vertex (PV), the impact parameter (IP), is measured with a resolution of $(15+29/\pt)\mum$, where \pt is the component of the momentum transverse to the beam, in\,\gevc. Different types of charged hadrons are distinguished using information from two ring-imaging Cherenkov detectors (\rich). Photons, electrons and hadrons are identified by a calorimeter system consisting of scintillating-pad and preshower detectors, an electromagnetic calorimeter and a hadronic calorimeter. Muons are identified by a system composed of alternating layers of iron and multiwire proportional chambers, and gas electron multiplier detectors. 

The online event selection is performed by a trigger~\cite{LHCb-DP-2012-004}, which consists of a hardware stage, based on information from the calorimeter and muon systems, followed by a software stage, which applies a full event reconstruction. Signal events considered in the analysis must fulfil hardware and software trigger requirements. At the hardware trigger stage, events are required to have a muon with high \pt or a hadron, photon or electron with high transverse energy in the calorimeters. At the software stage, at least one charged particle should have high \pt and large \chisqip with respect to any PV, where \chisqip is defined as the difference in the vertex-fit \chisq of a given PV fitted with and without the considered track. The software trigger designed to select \bquark-hadron decays uses a multivariate algorithm~\cite{BBDT} to identify a two-, three- or four-track secondary vertex with a large scalar sum of the \pt of the associated charged particles and a significant displacement from the PVs. The PVs are fitted with and without the \B candidate, and the PV with the smallest \chisqip is associated with the \B candidate.

The analysis presented is based on $pp$ collision data corresponding to an integrated luminosity of 1\invfb at a centre-of-mass energy of 7\tev collected in 2011, 2\invfb at 8\tev collected in 2012 (jointly referred to as Run 1), and 1.8\invfb at 13\tev collected in 2015 and 2016 (referred to as Run 2). There are several differences between data collected in Run 1 and Run 2. The main difference is the higher $b\bar{b}$ production cross-section in Run 2~\cite{LHCB-PAPER-2015-037}. The average number of $pp$ interactions per bunch crossing is reduced to 1.1 in Run 2 compared to 1.7 in Run 1. The net effect is that, despite the higher energy of the collisions, the background levels and signal-to-background ratios in Run 1 and Run 2 for the type of decay analysed here are similar. Before the start of Run 2, the aerogel radiator was removed from the first \rich detector~\cite{LHCb-DP-2012-003}, which improves the detector resolution. Hence, for momenta typical of decays in this analysis, the particle identification criteria have resulted in an increased efficiency of signal selection while simultaneously decreasing the rate of misidentified backgrounds. For the \decay{\Bm}{\D(\Km\pip)\Kstarm} decay mode, the combination of higher $b\bar{b}$ production cross-section, improved particle identification and improvements to the online selection in Run 2 have resulted in a factor of three increase in the yield for a given integrated luminosity.

Simulated event samples are used for the study of efficiencies. In the simulation, $pp$ collisions are generated using \pythia~\cite{Sjostrand:2006za,*Sjostrand:2007gs} with a specific \lhcb configuration~\cite{LHCb-PROC-2010-056}.  Decays of hadronic particles are described by \evtgen~\cite{Lange:2001uf}, in which final-state radiation is generated using \photos~\cite{Golonka:2005pn}. The interaction of the generated particles with the detector, and its response, are implemented using the \geant toolkit~\cite{Allison:2006ve, *Agostinelli:2002hh} as described in Ref.~\cite{LHCb-PROC-2011-006}.

\section{Offline selection}
\label{sec:selection}

The \Kstarm meson is reconstructed in the decay \hbox{\decay{\Kstarm}{\KS\pim}} and the \KS meson is reconstructed through its decay to two charged pions. If the pions from the \KS decay leave sufficient hits in the \velo to be included in the track reconstruction, the reconstructed \KS meson is called ``long". Due to the high boost from the $pp$ collision many \KS particles decay outside the \velo. If the pions from the \KS decay do not leave sufficient hits in the \velo, the reconstructed \KS meson is called ``downstream", with the first hits being recorded in the TT, which typically results in poorer mass resolution. These \KS reconstruction types are treated as separate data samples and a slightly different selection is applied to each.
 
Reconstructed \B candidates are formed by combining a \Kstarm candidate with a \D candidate, which are required to form a good-quality vertex.
For each \D, \Kstarm, and \KS candidate the reconstructed meson masses are required to lie within 25\mevcc around the \D mass, 75\mevcc around the \Kstarm mass, and 15\mevcc around the \KS mass for long candidates and 20\mevcc for downstream candidates~\cite{PDG2016}. A kinematic fit~\cite{Hulsbergen:2005pu} is performed on the full \B decay chain constraining the \B candidate to point towards the PV, and the \D and \KS candidates to have their known masses~\cite{PDG2016}. To suppress charmless backgrounds, the \D decay vertex is required to be well-separated from and downstream of the \Bm decay vertex. Also, the \KS decay vertex is required to be well-separated from and downstream of the \Bm decay vertex in order to suppress \decay{\Bm}{\D\pim\pip\pim} decays. The selection window of $\pm$75\mevcc, 1.5 times the $K^{*}(892)^-$ natural width, is required to suppress $\Bm \to D \KS \pim$ decays that do not proceed via an intermediate $K^{*}(892)^-$ resonance. Further suppression of these decays is achieved by requiring the magnitude of the cosine of the \KS helicity angle to be greater than 0.3. The \KS helicity angle is defined as the angle between the \KS and the \Bm momentum vectors in the \Kstarm rest frame. This requirement retains 97\% of true \Kstarm decays, which are distributed parabolically in this variable, while rejecting 30\% of the background.

Requirements, based mainly on the RICH system, are applied to all \D decay products to identify them as kaons or pions. These selections are applied such that each \D candidate is assigned a unique category. Cross-feed between the \Km\pip, \Kp\Km and \pip\pim\ \D final states is negligible because after misidentification of a \pim meson as a \Km meson (or vice versa) the reconstructed mass of the \D meson lies outside the \D mass selection window. However, the favoured decay \decay{\Bm}{\D(\Km\pip)\Kstarm} can appear in the \decay{\Bm}{\D(\pim\Kp)\Kstarm} sample due to misidentification of both \D decay products. To suppress this, a veto is applied to the ADS decay mode. The \D mass is reconstructed assuming the mass hypotheses of the decay products are swapped. If the resulting value is within 15\mevcc of the nominal \D mass, the candidate is removed from the sample, after which any remaining contamination is negligible while retaining 92\% of the signal. Similarly a 15\mevcc veto selection is applied to the four-body ADS decay mode to prevent the contamination of \decay{\Bm}{\D(\Km\pip\pim\pip)\Kstarm} in the \decay{\Bm}{\D(\pim\Kp\pim\pip)\Kstarm} sample. The swapped \D mass hypothesis is considered for both \pip mesons separately, resulting in a combined signal efficiency for the vetoes of 90\%.

Combinatorial background is suppressed using a Boosted Decision Tree (BDT) multivariate discriminant~\cite{Breiman}. To train the BDT for two-body decays, simulated \decay{\Bm}{\D(\Km\pip)\Kstarm} candidates are used as a signal sample and events from the high-mass sideband region of the \Bm mass, above 5600\mevcc, in the favoured \decay{\Bm}{\D(\Km\pip)\Kstarm} decay mode are used as a sample of combinatorial background. An analogous strategy is employed in the BDT for four-body decays. Various input quantities are used to exploit the topology of the decay; of particular importance are the \Bm vertex-fit $\chi^2$ and the \pt asymmetry between the \Bm candidate and other tracks from the same PV, defined as
\begin{equation}
A_{\pt} = \frac{p_{\rm T}^{B} - p_{\rm T}^{\text{cone}}}{p_{\rm T}^{B} + p_{\rm T}^{\text{cone}}}
\end{equation}
where $p_{\rm T}^{B}$ is the \pt of the reconstructed \Bm signal candidate and $p_{\rm T}^{\text{cone}}$ is the scalar sum of the \pt of all other tracks in a cone surrounding the \Bm candidate. This asymmetry is a quantitative measure of the isolation of the \Bm candidate. Other input quantities used include the logarithm of the \chisqip for various particles and the \pt of the \KS candidate (for downstream candidates only). The selection requirement on the BDT output was chosen to minimise the uncertainty on the \CP observables. The optimisation is performed separately for the GLW and ADS decay modes. Averaged across the whole dataset used for the analysis, the BDT selection applied to the favoured \decay{\Bm}{\D(\Km\pip)\Kstarm} channel gives a signal efficiency of 95\% (90\%) and a background rejection of 94\% (95\%) for long (downstream) candidates. Similarly, the four-body favoured \decay{\Bm}{\D(\Km\pip\pim\pip)\Kstarm} channel gives a signal efficiency of 95\% (93\%) and a background rejection of 96\% (97\%) for long (downstream) candidates.

\section{Fit to the invariant mass distribution}
\label{sec:cpfit}

\begin{figure}[h]
\centering
\includegraphics[width=\linewidth]{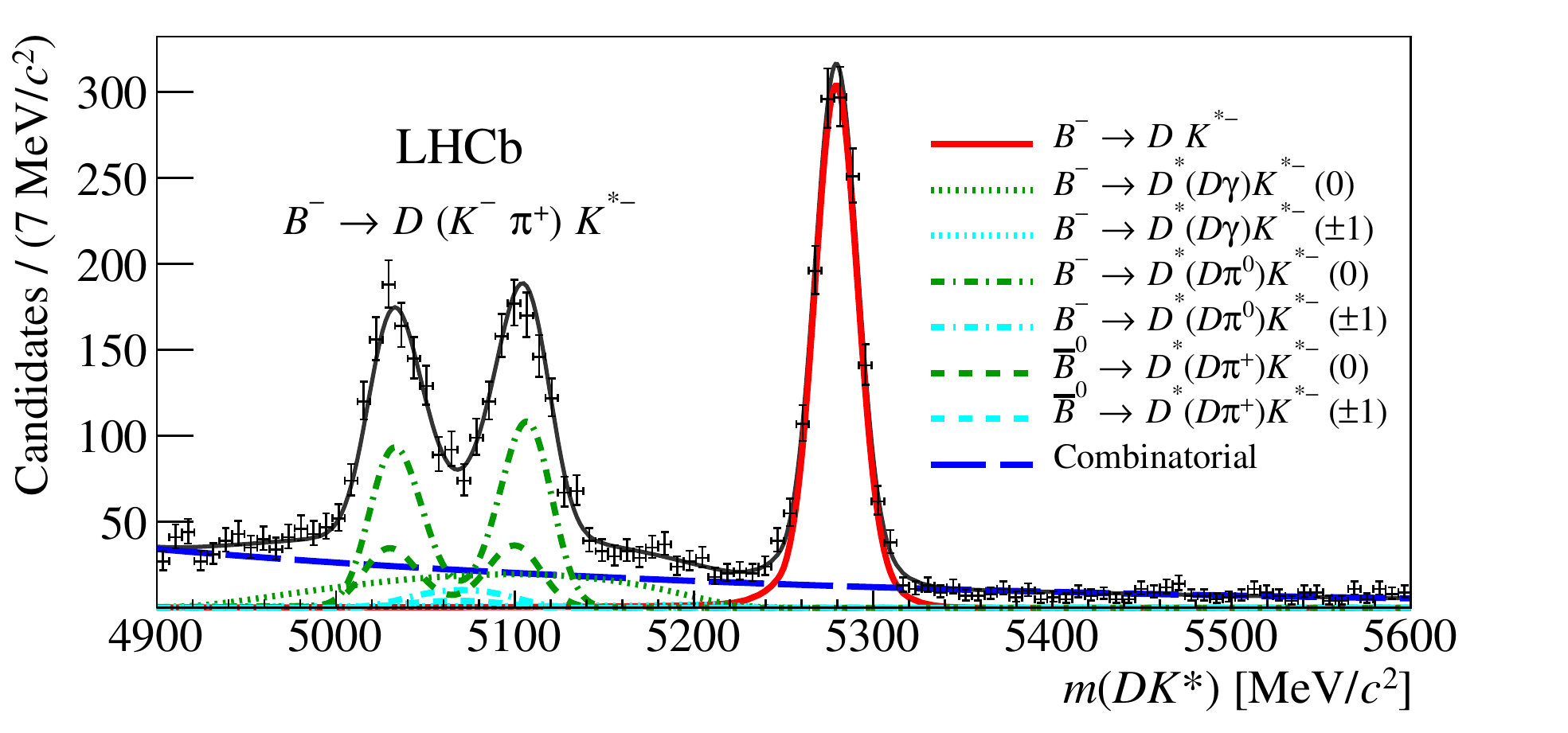}
\hfill
\includegraphics[width=\linewidth]{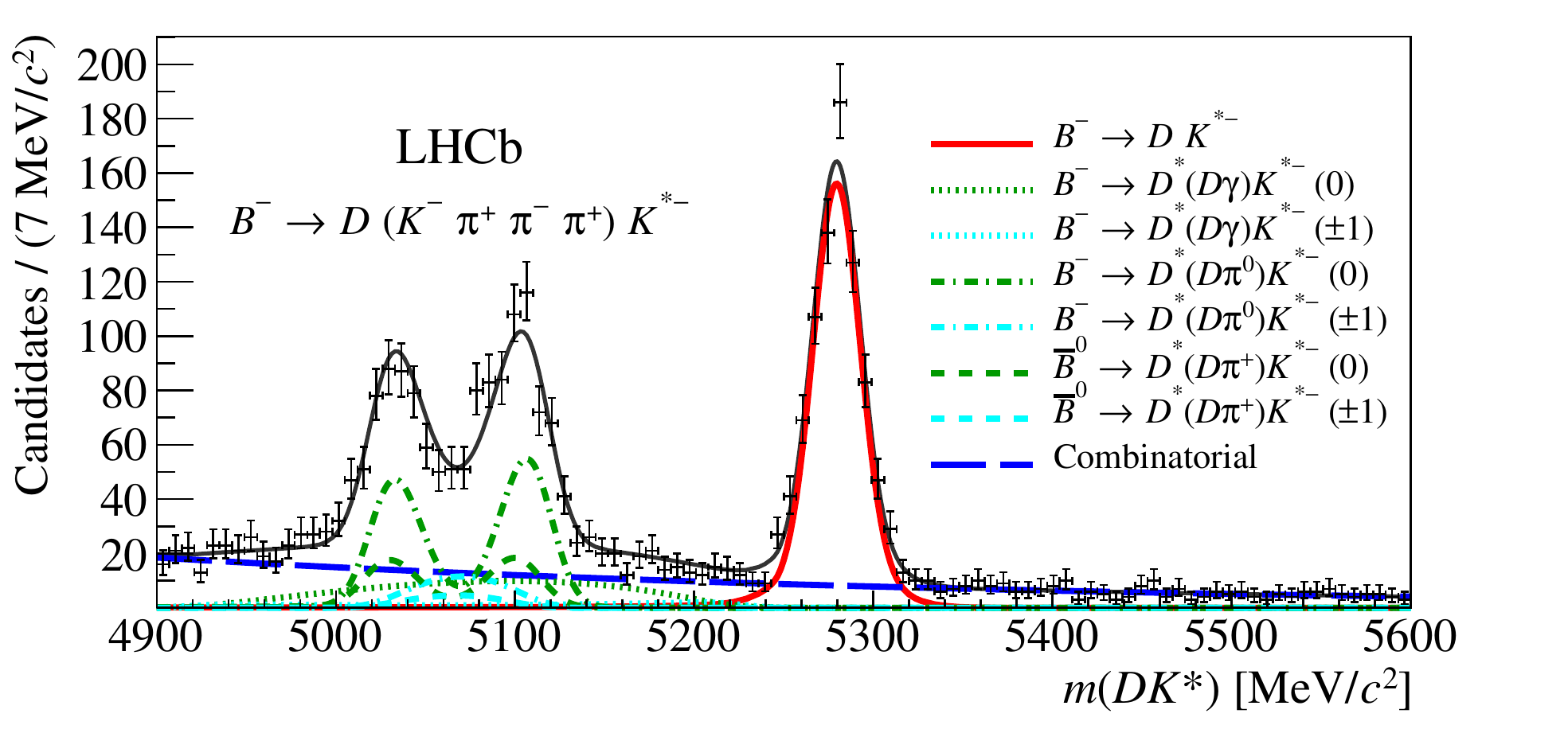}
\caption{Invariant mass distribution with the fit result superimposed for the favoured \mbox{\decay{\Bm}{\D(\Km\pip)\Kstarm}} decay mode (top), and \mbox{\decay{\Bm}{\D(\Km\pip\pim\pip)\Kstarm}} decay mode (bottom), using Run 1 and Run 2 data combined. The labels 0 and $\pm$1 correspond to the helicity state of the \Dstar meson.}
\label{massfit}
\end{figure}

Extended unbinned maximum likelihood fits are applied to the \B candidate mass spectra, in the mass range 4900--5600\mevcc, for candidates reconstructed in the favoured decay modes \decay{\Bm}{\D(\Km\pip)\Kstarm} and \decay{\Bm}{\D(\Km\pip\pim\pip)\Kstarm}. The same fit model is applied to both spectra. The model consists of a signal component, backgrounds from partially reconstructed decays and a combinatorial background shape. The charmless background has been suppressed to negligible levels, therefore no component is included in the fit. The signal component is described by the sum of two Crystal Ball (CB) functions~\cite{Skwarnicki:1986xj} with the same peak position, which contain small radiative tails that extend towards lower invariant mass. The signal shape parameters are determined from simulation, except for the common peak position and one of the widths, which are allowed to vary in the fit. The combinatorial background is described by an exponential function. The results of these fits are shown in Fig.~\ref{massfit}. 

Backgrounds from partially reconstructed decays include \decay{\B}{\Dstar\Kstar} decays where a pion or photon is not reconstructed, namely \hbox{\decay{\Bm}{\Dstarz(\Dz\piz)\Kstarm}}, \hbox{\decay{\Bm}{\Dstarz(\Dz\gamma)\Kstarm}} and \hbox{\decay{\Bdb}{\Dstarp(\Dz\pip)\Kstarm}}. These are decays of \B mesons into two vector particles, which are described by three independent helicity amplitudes, corresponding to the helicity states of the \Dstar meson, denoted by $-1$, $0$ and $+1$. The reconstructed \B-candidate mass distributions for $-1$ and $+1$ helicity states are indistinguishable so these states are collectively named $\pm1$. Therefore, for each \Dstar\Kstarm channel, two different components are considered, $0$ and $\pm1$. The shape of these components are determined from simulations and parameterised as Gaussian functions convolved with a second-order polynomial, described in detail in Ref.~\cite{LHCB-PAPER-2017-021,LHCB-PAPER-2016-006}, with all parameters fixed in the fit. The ratio between the yields of the three \Dstar\Kstarm decay modes are fixed according to their branching fractions and selection efficiencies, assuming no \CP violation. This procedure assumes that the longitudinal polarisation fraction for \Dstar\Kstarm decays is the same for \Bdb and \Bm mesons. The total partially reconstructed yield is allowed to vary as well as the yield ratio between the sum of the 0 shapes and the sum of the $\pm 1$ shapes.

As seen from the fit projections in Fig.~\ref{massfit}, these background contributions are sufficient to describe the overall invariant mass distribution of the favoured decay mode. A number of other backgrounds which could appear close to the signal peak are studied in simulation and found to be negligible, for example \decay{\Bm}{\D\Kstarm\piz} and \decay{\Bm}{\D(\KS\pi\pi)\Km}. Figure~\ref{massfit} shows that the main background contribution near the signal peak is combinatorial background, while only a small amount of partially reconstructed background enters the signal region. A significant fraction of the combinatorial background is expected to come from \decay{\Bm}{\D\pim X} decays combined with a real but unrelated \KS meson, which is consistent with the observed difference in background level between the \decay{\Bm}{\D(\Km\pip)\Kstarm} and \decay{\Bm}{\D(\Kp\pim)\Kstarm} decay modes. In the case of the \decay{\Bm}{\D(\Kp\Km)\Kstarm} decay mode, an additional background coming from the decay \decay{\Lb}{\Lc(p\Km\pip)\Kstarm} needs to be considered, where the \pip meson is not reconstructed and the proton is misidentified as a kaon. The shape of this background is obtained by parameterising the mass distribution from simulated background events; the shape parameters are fixed in the fits described below. The yield of \decay{\Lb}{\Lc(p\Km\pip)\Kstarm} compared to signal in the \decay{\Bm}{\D(\Km\pip)\Kstarm} favoured decay mode is allowed to vary.

Restricting the lower limit of the mass range to 5230\mevcc removes 0.4\% of signal and avoids the need to fit the backgrounds from partially reconstructed decays in each of the decay modes. This strategy improves fit stability in the decay modes with lower yields. The shape and yield of the small amount of background from partially reconstructed decays present in all \D decay categories above 5230\mevcc is determined and fixed from the fit to data with the favoured decay, adjusted for the smaller branching fractions of the rarer \D decays. The yield is estimated to be less than one candidate for all \CP-violating decay modes, and therefore uncertainties due to the assumptions present in the initial fit have a very small effect. These uncertainties in the yield, shape and possible asymmetries in the distribution between \Bp and \Bm are sources of systematic uncertainty.

A simultaneous fit is performed to 56 \B-meson mass distributions, corresponding to each of the seven \D decay modes (\Km\pip, \Kp\Km, \pip\pim, \Kp\pim, \Km\pip\pim\pip, \pip\pim\pip\pim and \Kp\pim\pip\pim), two \B-meson charges (\Bp and \Bm), two \KS reconstruction types (long and downstream) and two periods of data taking (Run 1 and Run 2). Based on fits to the data and simulation samples, the same signal peak position and width are used  for the two periods of data taking, \B-meson charges and \KS reconstruction types, but they are allowed to differ between two- and four-body decay modes. The combinatorial background slope is required to have the same value for all two- and four-body decay modes separately, but can differ between long and downstream categories.

The parameters determined from the simultaneous fit are the yields in the favoured signal decay modes and the \CP observables $A_{K\pi}$, $A_{KK}$, $A_{\pi\pi}$, $R_{KK}$, $R_{\pi\pi}$, $R^+_{K\pi}$, $R^-_{K\pi}$, $A_{K\pi\pi\pi}$, $A_{\pi\pi\pi\pi}$, $R_{\pi\pi\pi\pi}$, $R^+_{K\pi\pi\pi}$ and $R^-_{K\pi\pi\pi}$. The observables are related to the ratios between the yields through various efficiency corrections, given by

{\footnotesize
\begin{equation}
R_{hh} = \frac{N(\decay{\Bm}{D(h^+h^-)\Kstarm})}{N(\decay{\Bm}{D(\Km\pip)\Kstarm})} \times \frac{\BR(\decay{\Dz}{\Km\pip})}{\BR(\decay{\Dz}{hh})} \times \frac{\epsilon_{\text{sel}}(K\pi)}{\epsilon_{\text{sel}}(hh)} \times \frac{\epsilon_{\text{PID}}(K\pi)}{\epsilon_{\text{PID}}(hh)} \text{ ,}
\label{effcorrectionglw}
\end{equation}
\begin{equation}
R^{\pm}_{K\pi} = \frac{N(\decay{\Bpm}{D(\Kmp\pipm)\Kstarpm})}{N(\decay{\Bpm}{D(\Kpm\pimp)\Kstarpm})} \times \frac{\epsilon_{\text{sel}}(K\pi)}{\epsilon_{\text{sel}}(\pi K)} \times \frac{1}{\epsilon_{\text{veto}}(\pi K)} \text{ ,}
\label{effcorrectionads}
\end{equation}
\begin{equation}
R_{\pi\pi\pi\pi} = \frac{N(\decay{\Bm}{D(\pip\pim\pip\pim)\Kstarm})}{N(\decay{\Bm}{D(\Km\pip\pim\pip)\Kstarm})} \times \frac{\BR(\decay{\Dz}{\Km\pip\pim\pip})}{\BR(\decay{\Dz}{\pi\pi\pi\pi})} \times \frac{\epsilon_{\text{sel}}(K\pi\pi\pi)}{\epsilon_{\text{sel}}(\pi\pi\pi\pi)} \times \frac{\epsilon_{\text{PID}}(K\pi\pi\pi)}{\epsilon_{\text{PID}}(\pi\pi\pi\pi)} \text{ ,}
\label{effcorrectionglw4body}
\end{equation}
\begin{equation}
R^{\pm}_{K\pi\pi\pi} = \frac{N(\decay{\Bpm}{D(\Kmp\pipm\pimp\pipm)\Kstarpm})}{N(\decay{\Bpm}{D(\Kpm\pimp\pipm\pimp)\Kstarpm})} \times \frac{\epsilon_{\text{sel}}(K\pi\pi\pi)}{\epsilon_{\text{sel}}(\pi K\pi\pi)} \times \frac{1}{\epsilon_{\text{veto}}(\pi K\pi\pi)} \text{ ,}
\label{effcorrectionads4body}
\end{equation}}%
\noindent
where $\epsilon_{\text{sel}}$, $\epsilon_{\text{PID}}$ and $\epsilon_{\text{veto}}$ are the selection, particle-identification and veto efficiencies, respectively, $N$ is the yield of the specified decay and $h$ represents a \pion or \kaon meson. The veto is only applied to the ADS decay mode to reduce cross-feed from the favoured decay. These efficiencies are determined from simulation. The selection efficiency for various \D decay modes accounts for any differences in kinematics between these modes as well as a tighter BDT cut in the ADS decay mode, which is applied in order to optimise the uncertainty in the \CP observables. Any further correction to the four-body observables due to nonuniform acceptance was found to be negligible. The efficiencies cancel for the determination of the \CP asymmetries, while corrections are applied for the \Bp, \Bm production asymmetry, $A_{\text{prod}}$, and decay mode dependent detection asymmetries, $A_{\text{det}}$, which are taken from previous \lhcb measurements for production asymmetry~\cite{LHCb-PAPER-2016-054}, kaon detection asymmetry~\cite{LHCb-PAPER-2014-013} and pion detection asymmetry~\cite{LHCB-PAPER-2012-009}. The value $A_{\text{prod}}$ is assumed to be the same for 7\tev, 8\tev and 13\tev data. A possible difference in $A_{\text{prod}}$ for Run 2 data compared to Run 1 is accounted for as a systematic uncertainty. As the asymmetries are small, \order{(1\%)} or less, the observed uncorrected asymmetry $A_{\text{raw}}$ can be expressed as the sum $A_{\text{raw}} = A_{\text{phys}} + A_{\text{prod}} + A_{\text{det}}$, where $A_{\text{phys}}$ is the \CP asymmetry to be extracted. Hence, $A_{\text{prod}}$ and $A_{\text{det}}$ provide additive corrections to the measured asymmetry. 

\section{Results}
\label{sec:results}

\begin{figure}[!h]
\centering
\subfloat{\includegraphics[trim = 0mm 0mm 8mm 2mm,clip,width=0.9\linewidth]{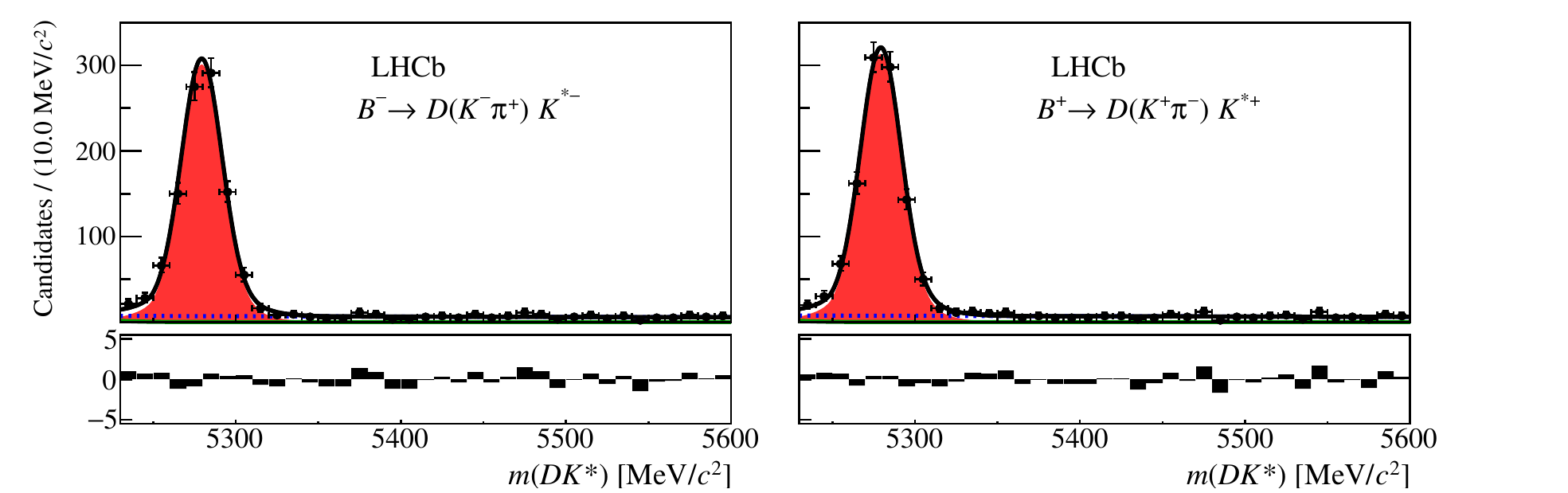}}
\hfill
\subfloat{\includegraphics[trim = 0mm 0mm 8mm 2mm,clip,width=0.9\linewidth]{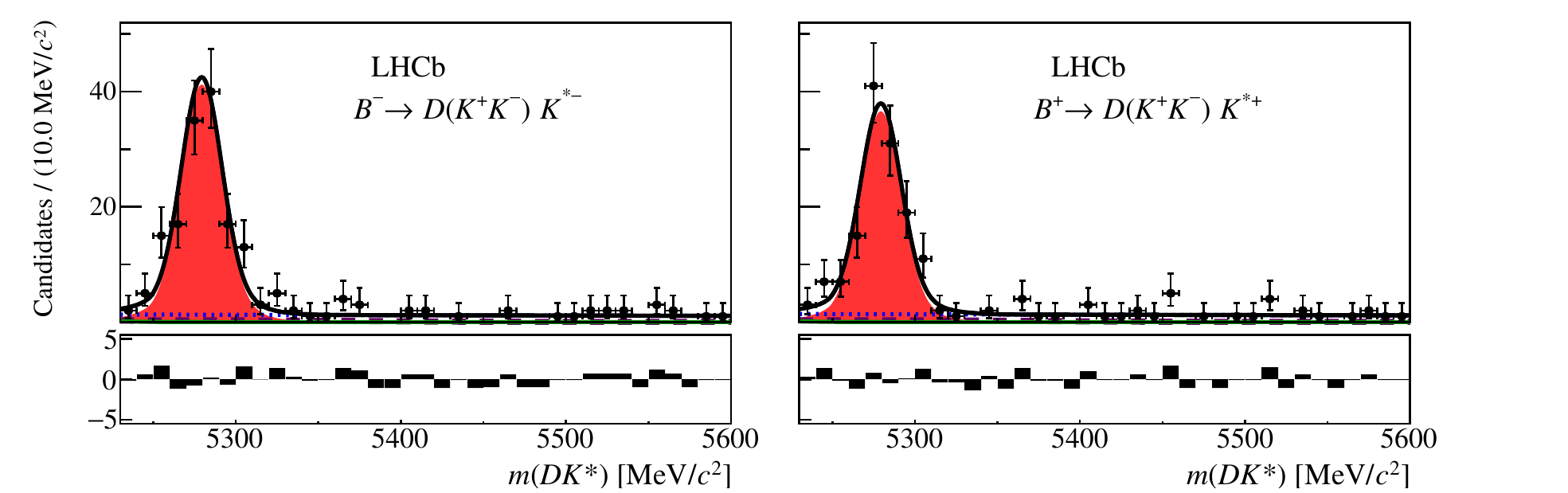}}
\hfill
\subfloat{\includegraphics[trim = 0mm 0mm 8mm 2mm,clip,width=0.9\linewidth]{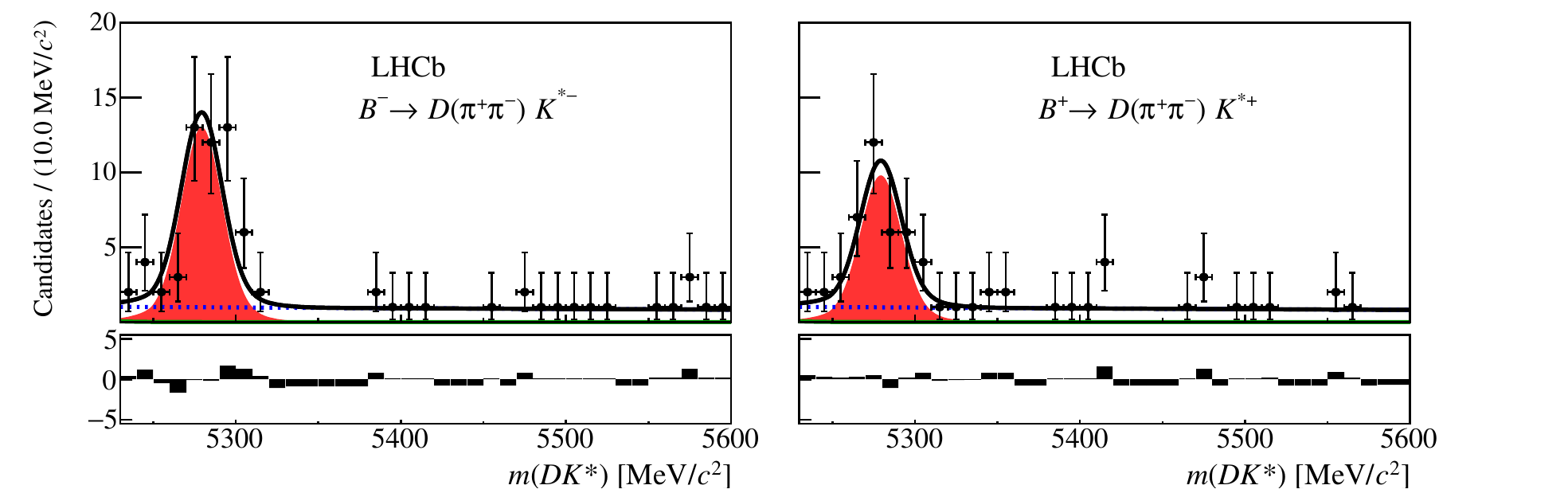}}
\hfill
\subfloat{\includegraphics[trim = 0mm 0mm 8mm 2mm,clip,width=0.9\linewidth]{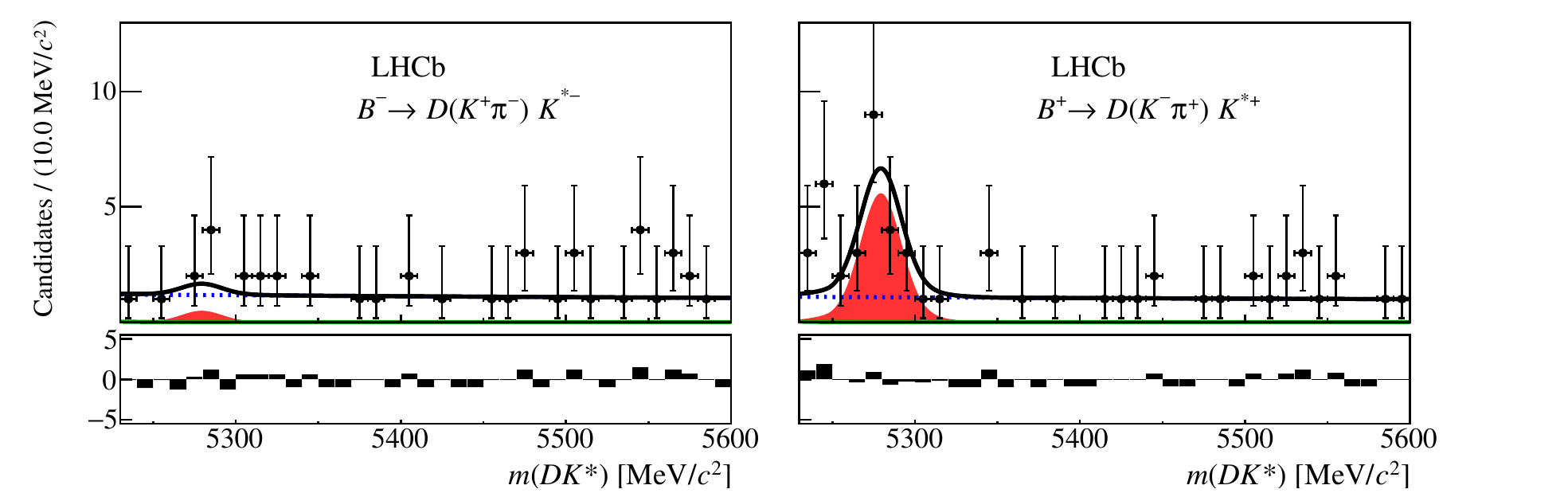}}
\caption{Result of fits to data for the two-body decay modes with Run 1, Run 2, long and downstream categories summed for presentation. The signal is represented by the red shaded area, the combinatorial background by the dotted blue line and the partially reconstructed background by the solid green line. In the \decay{\Dz}{\Kp\Km} fits the \decay{\Lb}{\Lc\Kstarm} background is represented by the dashed purple line. The total fit is given by the black line. The residuals, shown below each plot, are defined as the difference between the data and the fit value in each bin, normalised by the uncertainty.}
\label{dataplots2body}
\end{figure}

\begin{figure}[!h]
\vspace{75pt}
\centering
\subfloat{\includegraphics[trim = 0mm 0mm 8mm 2mm,clip,width=0.9\linewidth]{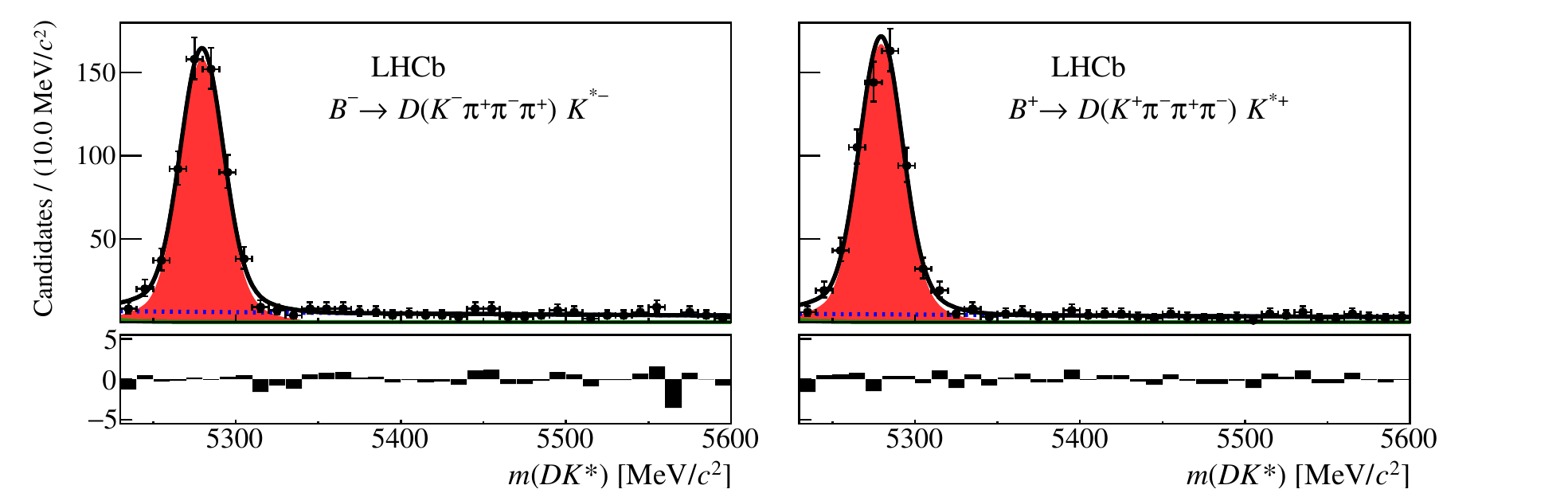}}
\hfill
\subfloat{\includegraphics[trim = 0mm 0mm 8mm 2mm,clip,width=0.9\linewidth]{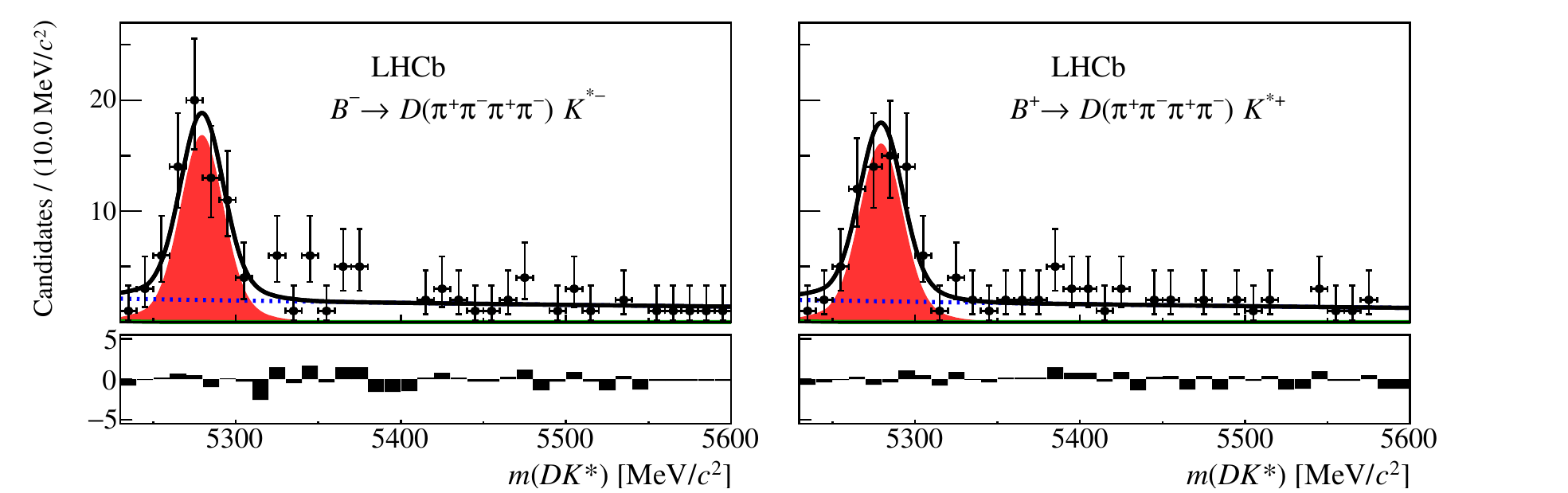}}
\hfill
\subfloat{\includegraphics[trim = 0mm 0mm 8mm 2mm,clip,width=0.9\linewidth]{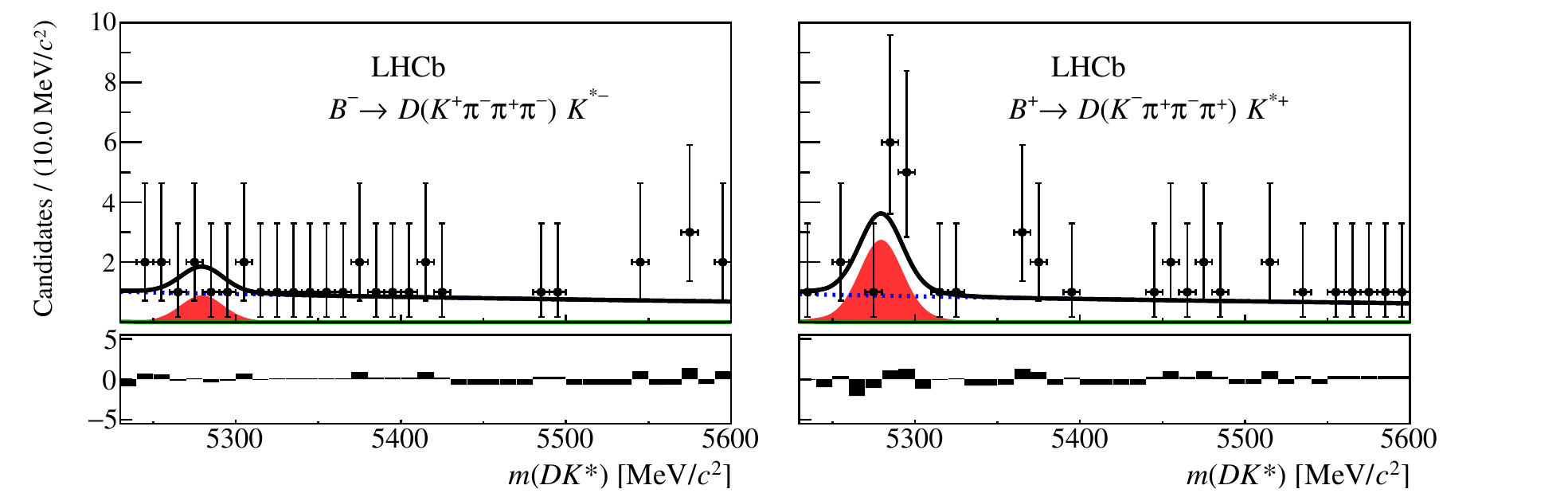}}
\caption{Result of fits to data for the four-body decay modes with Run 1, Run 2, long and downstream categories summed for presentation. The signal is represented by the red shaded area, the combinatorial background by the dotted blue line and the partially reconstructed background by the solid green line. The total fit is given by the black line. The residuals are shown below each plot.\vspace{75pt}}
\label{dataplots4body}
\end{figure}

The invariant mass spectra and resulting fits to data, combining Run 1, Run 2, long and downstream categories, are shown in Figs.~\ref{dataplots2body} and~\ref{dataplots4body}. The yields determined from the fitted parameters are given in Table~\ref{yields}. The Wilks' theorem statistical significance~\cite{Wilks:1938dza} for the two-body ADS decay mode is 4.2$\sigma$, while for the four-body ADS decay mode it is 2.8$\sigma$. This represents the first evidence of the two-body suppressed decay.

\begin{table}
\centering
\caption{Fitted yields in each of the \B decay modes. The uncertainties are statistical only.}
\begin{tabular}{c|cc}
\hline
Decay mode & \Bm yield & \Bp yield \\
\hline
$\begin{aligned}[t]
&\decay{\Bpm}{\D(\Kpm\pimp)\Kstarpm} \\
&\decay{\Bpm}{\D(\Kp\Km)\Kstarpm} \\
&\decay{\Bpm}{\D(\pip\pim)\Kstarpm} \\
&\decay{\Bpm}{\D(\Kmp\pipm)\Kstarpm} \\
&\decay{\Bpm}{\D(\Kpm\pimp\pip\pim)\Kstarpm} \\
&\decay{\Bpm}{\D(\pip\pim\pip\pim)\Kstarpm} \\
&\decay{\Bpm}{\D(\Kmp\pipm\pim\pip)\Kstarpm}\\
\end{aligned}$ & 
$\begin{aligned}[t]
996 &\pm 34 \\
134 &\pm 14 \\
45 &\pm 10 \\
1.6 &\pm 1.9 \\
556 &\pm 26 \\
59 &\pm 10 \\
3 &\pm 5 \\
\end{aligned}$ &
$\begin{aligned}[t]
1035 &\pm 35 \\
121 &\pm 13 \\
33 &\pm 9 \\
19 &\pm 7 \\
588 &\pm 27 \\
56 &\pm 10 \\
10 &\pm 6 \\
\end{aligned}$ \\
\hline
\end{tabular}
\label{yields}
\end{table}
%

Branching fractions~\cite{PDG2016}, various efficiencies and asymmetries are used as inputs to the simultaneous fit in order to relate the measured yields to the \CP observables. Each of these inputs has an associated uncertainty which needs to be propagated to the \CP observables giving rise to the systematic uncertainties. In the case of the efficiencies, uncertainties arise from a limited sample size of simulated events. Uncertainties on $A_{\text{prod}}$ and $A_{\text{det}}$ are taken from previous \lhcb measurements in Run 1~\cite{LHCb-PAPER-2016-054,LHCb-PAPER-2014-013,LHCB-PAPER-2012-009}. The changes to the detector between the data-taking periods are not expected to significantly affect the $A_{\text{det}}$ measurement. For $A_{\text{prod}}$, a conservative estimate, double the Run 1 uncertainty, is assigned to accommodate a possible dependence of the production asymmetry on the centre-of-mass energy. The systematic uncertainties due to the use of fixed inputs from branching ratios, simulation efficiencies, asymmetry corrections and shape parameters are estimated by performing multiple fits to data where each relevant parameter is varied according to a Gaussian distribution with the width as the assigned uncertainty. The standard deviation of the fitted parameter distribution is assigned as the systematic uncertainty. Correlations between the shape parameters are small, typically less than 10\%, and are ignored. Tests, where the most relevant correlations have been included, show a negligible impact on the systematic uncertainty arising from the fixed shape parameters. 

Other systematic uncertainties arise from the modelling of the signal and partially reconstructed backgrounds and the effect of any residual charmless \B decays. The systematic uncertainties from these sources are computed by generating pseudoexperiments. In each case the generated model is varied according to the systematic effects being estimated. The systematic uncertainty on each observable is taken to be the difference between the mean of the fitted parameter distribution and the generated value. The systematic uncertainty on the partially reconstructed background takes into account uncertainties in the yield and shape parameters, as well as possible asymmetries due to \CP violation. The contamination from charmless \B decays is consistent with zero, although it has a large uncertainty. Pseudoexperiments are generated with charmless decays according to the fit model, with the number of events fluctuating according to the uncertainty in the fit. The assumption that the slope of the function describing the combinatorial background is the same for all \D decay modes has an associated uncertainty. Pseudoexperiments are generated fixing the slope parameters to a different value for each decay mode, where the value used is obtained from fits in the mass region above the \B mass. For the ADS mode, a potential background from \decay{\Bs}{\D(\Kstar(1410)^0 \to \Kstar(892)^-\pip)}, where the \pip meson is not reconstructed, is considered. An estimate of the contribution using simulated events and the branching fraction~\cite{LHCB-PAPER-2014-036} is found to be $2.6 \pm 2.6$ events, which is consistent with observations from data in the region of \B mass below the lower limit of the simultaneous fit. The shape of this background is obtained by parameterising the mass distribution from simulated events. A systematic is assigned by performing many fits to data varying the yield according to a Gaussian distribution with the width as the assigned uncertainty. The standard deviation of the fitted parameter distribution is assigned as the systematic uncertainty. A summary of the components of the systematic uncertainties for the \CP observables is given in Table~\ref{systematics}.

\begin{sidewaystable}[htbp]
\caption{Summary of systematic uncertainties. Uncertainties are not shown if they are more than two orders of magnitude smaller than the statistical uncertainty.}
\centering
{\footnotesize
\begin{tabular}{ccccccccccccc}
\hline	
\rule{0pt}{2.5ex}\rule[-1.2ex]{0pt}{0ex} & $A_{K\pi}$ & $A_{KK}$ & $A_{\pi\pi}$ & $R_{KK}$ & $R_{\pi\pi}$ & $R^+_{K\pi}$ & $R^-_{K\pi}$ & $A_{K\pi\pi\pi}$ & $A_{\pi\pi\pi\pi}$ & $R_{\pi\pi\pi\pi}$ & $R^+_{K\pi\pi\pi}$ & $R^-_{K\pi\pi\pi}$ \\
\hline
Statistical & $0.023$ & $0.07$ & $0.13$ & $0.09$ & $0.15$ & $0.006$ & $0.004$ & $0.031$ & $0.11$ & $0.13$ & $0.008$ & $0.007$ \\
\hline
Branching fractions & $-$ & $-$ & $0.001$ & $0.013$ & $0.012$ & $-$ & $-$ & $-$ & $0.0008$ & $0.027$ & $-$ & $-$ \\
Selection efficiencies  & $-$ & $-$ & $-$ & $0.007$ & $0.006$ & $0.0002$ & $-$ & $-$ & $0.0008$ & $0.014$ & $-$ & $-$ \\
PID efficiencies  & $-$ & $-$ & $-$ & $0.002$ & $0.002$ & $-$ & $-$ & $-$ & $-$ & $0.002$ & $-$ & $-$ \\
Veto efficiencies  & $-$ & $-$ & $-$ & $-$ & $-$ & $0.0001$ & $-$ & $-$ & $-$ & $-$ & $-$ & $-$ \\
$A_{\text{prod}}$  & $0.0073$ & $0.007$ & $0.008$ & $-$ & $-$ & $-$ & $-$ & $0.0079$ & $0.0077$ & $-$ & $-$ & $-$ \\
$A_{\text{det}}$  & $0.0034$ & $0.003$ & $0.003$ & $-$ & $-$ & $0.0001$ & $-$ & $0.0034$ & $0.0030$ & $-$ & $0.0001$ & $-$ \\
Signal shape & $0.0011$ & $0.003$ & $0.003$ & $0.011$ & $0.027$ & $0.0011$ & $0.0013$ & $0.0017$ & $0.0022$ & $0.010$ & $0.0030$ & $0.0038$ \\
Combinatorial shape  & $0.0012$ & $0.003$ & $0.005$ & $0.004$ & $0.009$ & $0.0002$ & $0.0003$ & $0.0001$ & $0.0018$ & $-$ & $0.0012$ & $0.0004$ \\
Partially reconstructed shape  & $0.0007$ & $0.001$ & $0.003$ & $0.001$ & $0.005$ & $-$ & $0.0003$ & $0.0003$ & $0.0005$ & $0.002$ & $0.0008$ & $0.0001$ \\
Charmless  & $0.0008$ & $-$ & $0.003$ & $0.002$ & $0.007$ & $-$ & $0.0003$ & $0.0009$ & $0.0030$ & $0.002$ & $0.0008$ & $0.0001$ \\
\decay{\Lb}{\Lc\Kstarm} & $0.0002$ & $-$ & $-$ & $0.011$ & $0.001$ & $0.0001$ & $-$ & $-$ & $-$ & $-$ & $-$ & $-$ \\
\decay{\Bs}{\D\Kstar(1410)^0} & $-$ & $-$ & $-$ & $-$ & $-$ & $0.0005$ & $0.0001$ & $-$ & $-$ & $-$ & $-$ & $-$ \\
\hline
Total systematic & $0.0083$ & $0.009$ & $0.012$ & $0.022$ & $0.032$ & $0.0012$ & $0.0014$ & $0.0088$ & $0.0093$ & $0.032$ & $0.0034$ & $0.0038$ \\
\hline
\end{tabular}}
\label{systematics}
\end{sidewaystable}

The \CP observables determined from the fit shown in Figs.~\ref{dataplots2body} and~\ref{dataplots4body} are  
\begin{alignat*}{13}
A_{K\pi} &= &\ -&0.004&\ &\pm&\ &0.023&\ &\pm&\ &0.008& \\
A_{KK} &= &&0.06&\ &\pm&\ &0.07&\ &\pm&\ &0.01& \\
A_{\pi\pi} &= &&0.15&\ &\pm&\ &0.13&\ &\pm&\ &0.01& \\
R_{KK} &= &&1.22&\ &\pm&\ &0.09&\ &\pm&\ &0.02& \\
R_{\pi\pi} &= &&1.08&\ &\pm&\ &0.14&\ &\pm&\ &0.03& \\
R^+_{K\pi} &= &&0.020&\ &\pm&\ &0.006&\ &\pm&\ &0.001& \\ 
R^-_{K\pi} &= &&0.002&\ &\pm&\ &0.004&\ &\pm&\ &0.001& \\
A_{K\pi\pi\pi} &= &\ -&0.013&\ &\pm&\ &0.031&\ &\pm&\ &0.009& \\
A_{\pi\pi\pi\pi} &= &&0.02&\ &\pm&\ &0.11&\ &\pm&\ &0.01& \\
R_{\pi\pi\pi\pi} &= &&1.08&\ &\pm&\ &0.13&\ &\pm&\ &0.03& \\
R^+_{K\pi\pi\pi} &= &&0.016&\ &\pm&\ &0.007&\ &\pm&\ &0.003& \\ 
R^-_{K\pi\pi\pi} &= &&0.006&\ &\pm&\ &0.006&\ &\pm&\ &0.004&
\end{alignat*}
where the first uncertainty is statistical and the second is systematic. The correlation matrices for the statistical and systematic uncertainties are given in Tables~\ref{statisticalcorrelations} and \ref{systematiccorrelations}, respectively. The large correlations of the systematic uncertainties are mainly due to contributions from production and detection asymmetries. Combined results from the \Kp\Km and \pip\pim decay modes, taking correlations into account, are
\begin{alignat*}{13}
R_{\CP+} &= &\ &1.18&\ &\pm&\ &0.08&\ &\pm&\ &0.02& \\
A_{\CP+} &= &\ &0.08&\ &\pm&\ &0.06&\ &\pm&\ &0.01&
\end{alignat*}
where the first uncertainty is statistical and the second is systematic. In addition, $R^+$ and $R^-$ for the \Kp\pim and \Kp\pim\pip\pim decay modes can be transformed into the more commonly used $R_{ADS} = \left(R^- + R^+\right)/2\ $and \mbox{$A_{ADS} = \left(R^- - R^+\right)/\left(R^- + R^+\right)$}. These results, taking correlations into account, are
\begin{alignat*}{13}
R_{ADS}^{K\pi} &= &\ &0.011&\ &\pm&\ &0.004&\ &\pm&\ &0.001& \\
A_{ADS}^{K\pi} &= &\ -&0.81&\ &\pm&\ &0.17&\ &\pm&\ &0.04& \\
R_{ADS}^{K\pi\pi\pi} &= &\ &0.011&\ &\pm&\ &0.005&\ &\pm&\ &0.003& \\
A_{ADS}^{K\pi\pi\pi} &= &\ -&0.45&\ &\pm&\ &0.21&\ &\pm&\ &0.14&
\end{alignat*}
where the first uncertainty is statistical and the second is systematic. The measured asymmetries and ratios for the two-body \D meson decay modes are consistent with, and more precise than, the previous measurements from \babar~\cite{BaBarDKstar}.

\begin{table}[htbp]
\centering
\caption{Correlation matrix of the statistical uncertainties for the twelve physics observables from the simultaneous fit to data. Only half of the symmetric matrix is shown.}
{\scriptsize
\resizebox{\textwidth}{!}{
\begin{tabular}{c|cccccccccccc} 
\hline 
\rule{0pt}{2.5ex}\rule[-1.2ex]{0pt}{0ex}& $A_{K\pi}$ & $A_{KK}$ & $A_{\pi\pi}$ & $R_{KK}$ & $R_{\pi\pi}$ & $R^+_{K\pi}$ & $R^-_{K\pi}$ & $A_{K\pi\pi\pi}$ & $A_{\pi\pi\pi\pi}$ & $R_{\pi\pi\pi\pi}$ & $R^+_{K\pi\pi\pi}$ & $R^-_{K\pi\pi\pi}$ \\ 
 \hline
$A_{K\pi}$ & 1 & $-$ & $-$ & $-$ & $-$ & 0.08 & $-$0.01{\color{white}$-$} & $-$ & $-$ & $-$ & $-$ & $-$ \\
$A_{KK}$ & & 1 & $-$ & $-$ & $-$ & $-$ & $-$ & $-$ & $-$ & $-$ & $-$ & $-$ \\
$A_{\pi\pi}$ & & & 1 & $-$ & $-$0.02{\color{white}$-$} & $-$ & $-$ & $-$ & $-$ & $-$ & $-$ & $-$ \\
$R_{KK}$ & & & & 1 & 0.05 & 0.02 & $-$0.01{\color{white}$-$} & $-$ & $-$ & $-$ & $-$ & $-$ \\
$R_{\pi\pi}$ & & & & & 1 & 0.03 & 0.02 & $-$ & $-$ & $-$ & $-$ & $-$ \\
$R^+_{K\pi}$ & & & & & & 1 & 0.02 & $-$ & $-$ & $-$ & $-$ & $-$ \\
$R^-_{K\pi}$ & & & & & & & 1 & $-$ & $-$ & $-$ & $-$ & $-$ \\
$A_{K\pi\pi\pi}$ & & & & & & & & 1 & $-$ & $-$ & 0.07 & $-$0.03{\color{white}$-$} \\
$A_{\pi\pi\pi\pi}$ & & & & & & & & & 1 & 0.01 & $-$ & $-$ \\
$R_{\pi\pi\pi\pi}$ & & & & & & & & & & 1 & 0.04 & 0.04 \\
$R^+_{K\pi\pi\pi}$ & & & & & & & & & & & 1 & 0.03 \\
\rule[-1.2ex]{0pt}{0ex}$R^-_{K\pi\pi\pi}$ & & & & & & & & & & & & 1 \\
\hline 
\end{tabular}}}
\label{statisticalcorrelations}
\end{table}

\begin{table}[htbp]
\centering
\caption{Correlation matrix of the systematic uncertainties for the twelve physics observables from the simultaneous fit to data. Only half of the symmetric matrix is shown.}
{\scriptsize
\resizebox{\textwidth}{!}{
\begin{tabular}{c|cccccccccccc} 
\hline 
\rule{0pt}{2.5ex}\rule[-1.2ex]{0pt}{0ex}& $A_{K\pi}$ & $A_{KK}$ & $A_{\pi\pi}$ & $R_{KK}$ & $R_{\pi\pi}$ & $R^+_{K\pi}$ & $R^-_{K\pi}$ & $A_{K\pi\pi\pi}$ & $A_{\pi\pi\pi\pi}$ & $R_{\pi\pi\pi\pi}$ & $R^+_{K\pi\pi\pi}$ & $R^-_{K\pi\pi\pi}$ \\ 
 \hline
$A_{K\pi}$ & 1 & 0.82 & 0.72 & $-$ & $-$ & 0.01 & $-$0.02{\color{white}$-$} & 0.94 & 0.84 & $-$ & $-$0.01{\color{white}$-$} & $-$ \\
$A_{KK}$ & & 1 & 0.65 & $-$0.04{\color{white}$-$} & 0.02 & 0.01 & $-$0.02{\color{white}$-$} & 0.83 & 0.77 & $-$ & $-$ & $-$\\
$A_{\pi\pi}$ & & & 1 & $-$ & $-$0.03{\color{white}$-$} & $-$ & $-$0.02{\color{white}$-$} & 0.72 & 0.68 & $-$ & $-$ & 0.01 \\
$R_{KK}$ & & & & 1 & $-$ & 0.05 & 0.03 & $-$0.01{\color{white}$-$} & $-$ & $-$0.01{\color{white}$-$} & $-$0.01{\color{white}$-$} & $-$0.01{\color{white}$-$} \\
$R_{\pi\pi}$ & & & & & 1 & 0.06 & 0.08 & $-$0.01{\color{white}$-$} & $-$ & $-$0.01{\color{white}$-$} & $-$0.02{\color{white}$-$} & 0.01 \\
$R^+_{K\pi}$ & & & & & & 1 & 0.08 & $-$0.01{\color{white}$-$} & $-$ & $-$ & $-$0.01{\color{white}$-$} & $-$0.01{\color{white}$-$} \\
$R^-_{K\pi}$ & & & & & &  & 1 & $-$0.01{\color{white}$-$} & $-$0.01{\color{white}$-$} & $-$0.01{\color{white}$-$} & 0.01 & 0.03 \\
$A_{K\pi\pi\pi}$ & & & & & & & & 1 & 0.84 & $-$ & $-$0.01{\color{white}$-$} & $-$0.02{\color{white}$-$} \\
$A_{\pi\pi\pi\pi}$ & & & & & & & & & 1 & 0.03 & 0.01 & $-$ \\
$R_{\pi\pi\pi\pi}$ & & & & & & & & & & 1 & 0.01 & $-$0.01{\color{white}$-$} \\
$R^+_{K\pi\pi\pi}$ & & & & & & & & & & & 1 & 0.05 \\
\rule[-1.2ex]{0pt}{0ex}$R^-_{K\pi\pi\pi}$ & & & & & & & & & & & & 1 \\
\hline 
\end{tabular}}}
\label{systematiccorrelations}
\end{table}

\newpage
\section{Interpretation}
\label{sec:interpretation}

The \CP observables measured in this analysis can be used to determine the physics parameters $r_B$, $\delta_B$ and \Pgamma, via Eqs.~\ref{exp_Acp}-\ref{exp_Rpm4body}. The parameter $\kappa$ is estimated by generating many amplitude models for \decay{\B}{\D\KS\pi} decays~\cite{Laura} consisting of various resonant components whose relative amplitudes and phases are varied within limits according to the existing branching fraction measurements. The components used in the model are $\decay{\Bm}{\Dz K^*(892)^{-}}$ and the LASS lineshape~\cite{LASS}. The LASS lineshape is used to describe the $K\pi$ S-wave, which includes a nonresonant term and the $K_0^*(1430)^{-}$ resonance. Contributions from other resonances \eg \hbox{$\decay{K^*(1680)^-}{\KS\pim}$} and \hbox{$\decay{D_2^*(2460)^-}{\D\pim}$}, are considered to be negligible in the selected \Kstarm region and are not included in the model. For each model, the value of $\kappa$ is determined in the region of phase space defined by the \Kstarm mass window and \KS helicity angle requirements. The mean of the resulting distribution gives an estimate for $\kappa$ of $0.95 \pm 0.06$. The parameters $r_D^{K\pi}$, $\delta_D^{K\pi}$, $r_D^{K3\pi}$, $\delta_D^{K3\pi}$, $\kappa_{K3\pi}$ and $F_{4\pi}$ are also required as external inputs and are taken from Ref.~\cite{HFLAV16,charmk3pi,LHCB-PAPER-2015-057,charm4pi}. 

\begin{figure}[htbp]
\includegraphics[width=0.46\linewidth]{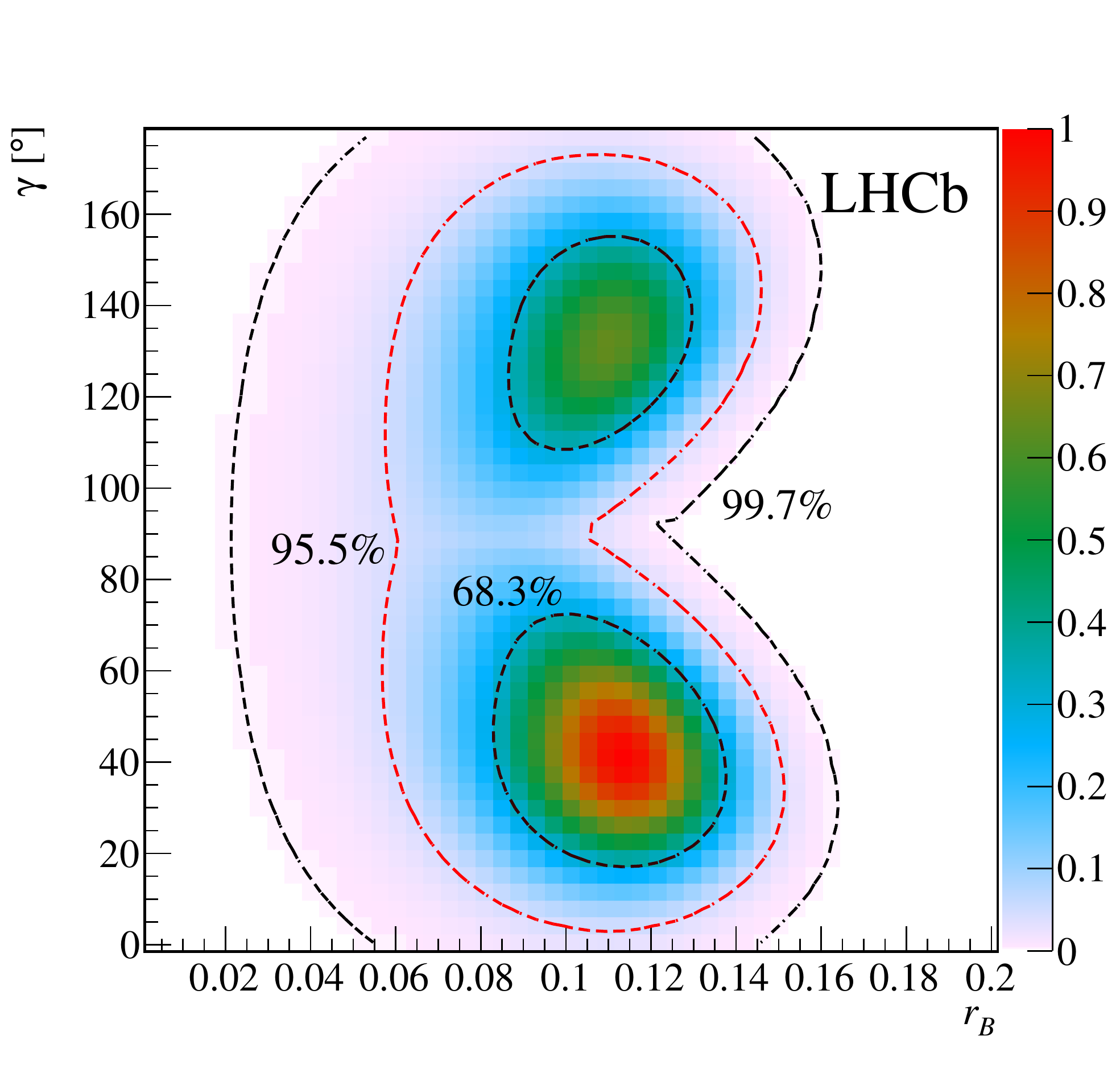}
\hfill
\includegraphics[width=0.46\linewidth]{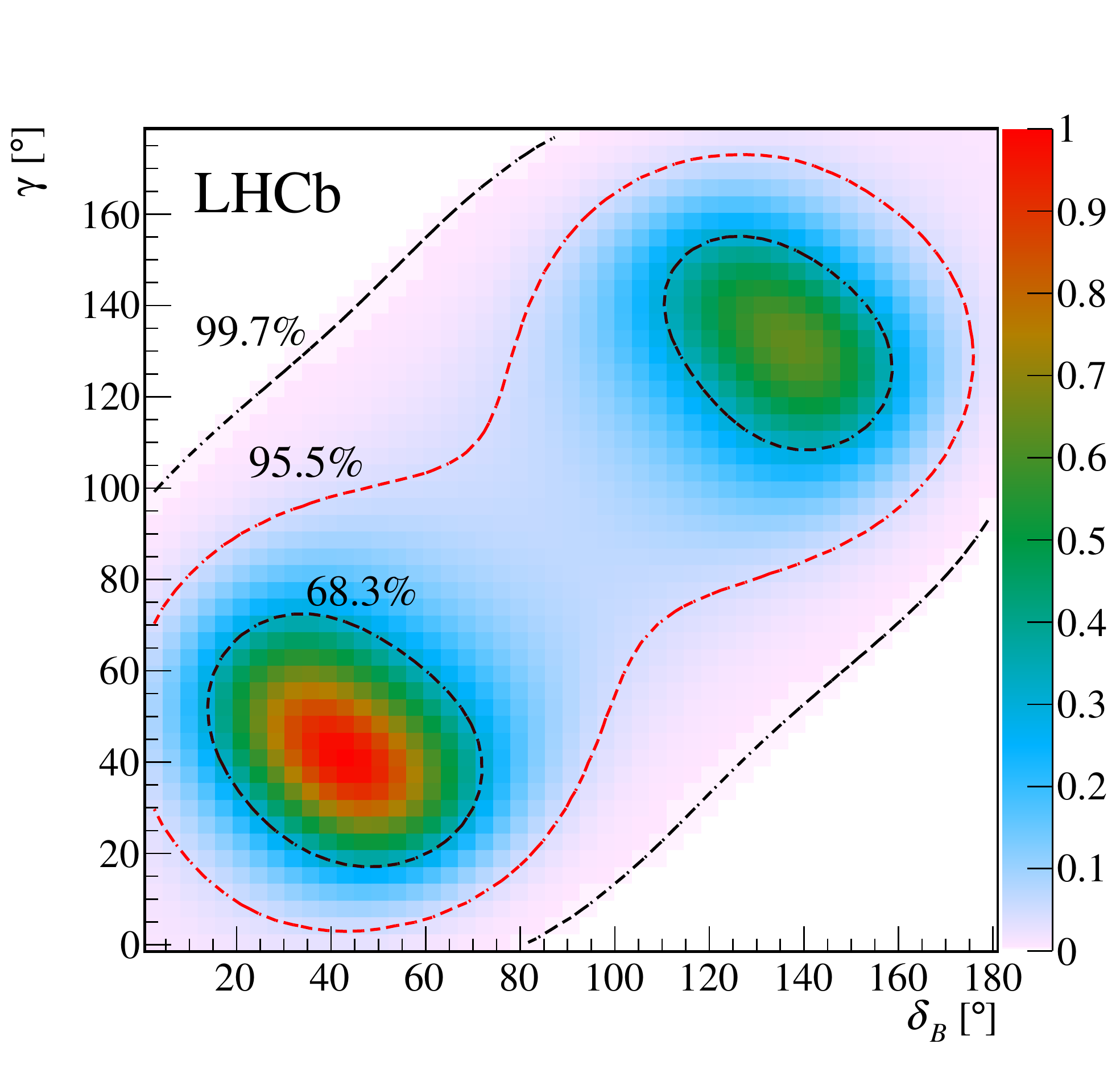}
\caption{Contour plots showing 2D scans of physics parameters $\gamma$ versus $r_B$ (left) and $\gamma$ versus $\delta_B$ (right). The dashed lines represent the $\Delta \chi^2 = 2.30,\ 6.18,\ 11.8$ contours, corresponding to 68.3\%, 95.5\%, 99.7\% confidence levels (CL), respectively. The colour scale represents $1-\text{CL}$.}
\label{gammadiniplots}
\end{figure}

Using the measured values of the \CP observables, their uncertainties and the covariance matrices, a global $\chi^2$ minimisation is performed, resulting in a minimum $\chi^2$ of 3.0 with 9 degrees of freedom. A scan of physics parameters is performed for a range of values and the difference in $\chi^2$ between the parameter scan values and the global minimum, $\Delta\chi^2$, is evaluated. The confidence level for any pair of parameters is calculated assuming that these are normally distributed, which enables the $\Delta \chi^2 = 2.30,\ 6.18,\ 11.8$ contours to be drawn, corresponding to 68.3\%, 95.5\%, 99.7\% confidence levels, respectively. These are shown in  Fig. \ref{gammadiniplots}. The data are consistent with the value of $\gamma$ indicated by previous measurements~\cite{LHCB-PAPER-2016-032, CKMfitter}, $\sim 70^\circ$, and result in a value of $r_B = 0.11 \pm 0.02$. This value of $r_B$ is determined at the point where the global $\chi^2$ of the fit is minimised.

\section{Conclusions}

A study of the \decay{\Bm}{\D\Kstarm} decay mode is presented where the \D meson decays to two- and four-body final states consisting of charged kaons and/or pions. The \CP observables $R_{CP+}$, $A_{CP+}$, $R^+_{K\pi}$, $R^-_{K\pi}$, $R_{\pi\pi\pi\pi}$, $A_{\pi\pi\pi\pi}$, $R^+_{K\pi\pi\pi}$ and $R^-_{K\pi\pi\pi}$ are measured from the high purity sample obtained from $pp$ collision data recorded with the \lhcb detector in Run 1 and Run 2. The measurement of the \CP asymmetries in the two-body decay modes and their ratio to the favoured decay mode is consistent with and more precise than the previous determination~\cite{BaBarDKstar}. While no bounds on \Pgamma are quoted due to the limited sensitivity of this decay mode in isolation, \decay{\Bm}{\D\Kstarm} decays will become valuable in constraining \Pgamma in the future, as more data are collected and more \D decay modes are analysed.
\clearpage
\section*{Acknowledgements}
%
%
\noindent We express our gratitude to our colleagues in the CERN
accelerator departments for the excellent performance of the LHC. We
thank the technical and administrative staff at the LHCb
institutes. We acknowledge support from CERN and from the national
agencies: CAPES, CNPq, FAPERJ and FINEP (Brazil); MOST and NSFC
(China); CNRS/IN2P3 (France); BMBF, DFG and MPG (Germany); INFN
(Italy); NWO (The Netherlands); MNiSW and NCN (Poland); MEN/IFA
(Romania); MinES and FASO (Russia); MinECo (Spain); SNSF and SER
(Switzerland); NASU (Ukraine); STFC (United Kingdom); NSF (USA).  We
acknowledge the computing resources that are provided by CERN, IN2P3
(France), KIT and DESY (Germany), INFN (Italy), SURF (The
Netherlands), PIC (Spain), GridPP (United Kingdom), RRCKI and Yandex
LLC (Russia), CSCS (Switzerland), IFIN-HH (Romania), CBPF (Brazil),
PL-GRID (Poland) and OSC (USA). We are indebted to the communities
behind the multiple open-source software packages on which we depend.
Individual groups or members have received support from AvH Foundation
(Germany), EPLANET, Marie Sk\l{}odowska-Curie Actions and ERC
(European Union), ANR, Labex P2IO, ENIGMASS and OCEVU, and R\'{e}gion
Auvergne-Rh\^{o}ne-Alpes (France), RFBR and Yandex LLC (Russia), GVA,
XuntaGal and GENCAT (Spain), Herchel Smith Fund, the Royal Society,
the English-Speaking Union and the Leverhulme Trust (United Kingdom).

\addcontentsline{toc}{section}{References}
\setboolean{inbibliography}{true}
\bibliographystyle{LHCb}
\bibliography{main}

\ifx\mcitethebibliography\mciteundefinedmacro
\PackageError{LHCb.bst}{mciteplus.sty has not been loaded}
{This bibstyle requires the use of the mciteplus package.}\fi
\providecommand{\href}[2]{#2}
\begin{mcitethebibliography}{10}
\mciteSetBstSublistMode{n}
\mciteSetBstMaxWidthForm{subitem}{\alph{mcitesubitemcount})}
\mciteSetBstSublistLabelBeginEnd{\mcitemaxwidthsubitemform\space}
{\relax}{\relax}

\bibitem{Cabibbo}
N.~Cabibbo, \ifthenelse{\boolean{articletitles}}{\emph{Unitary symmetry and
  leptonic decays},
  }{}\href{http://dx.doi.org/10.1103/PhysRevLett.10.531}{Phys.\ Rev.\ Lett.\
  \textbf{10} (1963) 531}\relax
\mciteBstWouldAddEndPuncttrue
\mciteSetBstMidEndSepPunct{\mcitedefaultmidpunct}
{\mcitedefaultendpunct}{\mcitedefaultseppunct}\relax
\EndOfBibitem
\bibitem{KM}
M.~Kobayashi and T.~Maskawa, \ifthenelse{\boolean{articletitles}}{\emph{{\CP
  violation in the renormalizable theory of weak interaction}},
  }{}\href{http://dx.doi.org/10.1143/PTP.49.652}{Progress of Theoretical
  Physics \textbf{49} (1973) 652}\relax
\mciteBstWouldAddEndPuncttrue
\mciteSetBstMidEndSepPunct{\mcitedefaultmidpunct}
{\mcitedefaultendpunct}{\mcitedefaultseppunct}\relax
\EndOfBibitem
\bibitem{CKMtriangle}
C.~Jarlskog, \ifthenelse{\boolean{articletitles}}{\emph{{Commutator of the
  quark mass matrices in the standard electroweak model and a measure of
  maximal \CP nonconservation}},
  }{}\href{http://dx.doi.org/10.1103/PhysRevLett.55.1039}{Phys.\ Rev.\ Lett.\
  \textbf{55} (1985) 1039}\relax
\mciteBstWouldAddEndPuncttrue
\mciteSetBstMidEndSepPunct{\mcitedefaultmidpunct}
{\mcitedefaultendpunct}{\mcitedefaultseppunct}\relax
\EndOfBibitem
\bibitem{LHCB-PAPER-2016-032}
LHCb collaboration, R.~Aaij {\em et~al.},
  \ifthenelse{\boolean{articletitles}}{\emph{{Measurement of the CKM angle
  $\gamma$ from a combination of LHCb results}},
  }{}\href{http://dx.doi.org/10.1007/JHEP12(2016)087}{JHEP \textbf{12} (2016)
  087}, \href{http://arxiv.org/abs/1611.03076}{{\normalfont\ttfamily
  arXiv:1611.03076}}\relax
\mciteBstWouldAddEndPuncttrue
\mciteSetBstMidEndSepPunct{\mcitedefaultmidpunct}
{\mcitedefaultendpunct}{\mcitedefaultseppunct}\relax
\EndOfBibitem
\bibitem{CKMfitter}
CKMfitter group, J.~Charles {\em et~al.},
  \ifthenelse{\boolean{articletitles}}{\emph{{CP violation and the CKM matrix:
  Assessing the impact of the asymmetric $B$ factories}},
  }{}\href{http://dx.doi.org/10.1140/epjc/s2005-02169-1}{Eur.\ Phys.\ J.\
  \textbf{C41} (2005) 1},
  \href{http://arxiv.org/abs/hep-ph/0406184}{{\normalfont\ttfamily
  arXiv:hep-ph/0406184}}, {updated results and plots available at:
  \href{http://ckmfitter.in2p3.fr}{{\tt http://ckmfitter.in2p3.fr}}}\relax
\mciteBstWouldAddEndPuncttrue
\mciteSetBstMidEndSepPunct{\mcitedefaultmidpunct}
{\mcitedefaultendpunct}{\mcitedefaultseppunct}\relax
\EndOfBibitem
\bibitem{LHCB-PAPER-2016-003}
LHCb collaboration, R.~Aaij {\em et~al.},
  \ifthenelse{\boolean{articletitles}}{\emph{{Measurement of $\CP$ observables
  in $\Bpm\to\D \Kpm$ and $\Bpm\to\D\pipm$ with two- and four-body $\D$
  decays}}, }{}\href{http://dx.doi.org/10.1016/j.physletb.2016.06.022}{Phys.\
  Lett.\  \textbf{B760} (2016) 117},
  \href{http://arxiv.org/abs/1603.08993}{{\normalfont\ttfamily
  arXiv:1603.08993}}\relax
\mciteBstWouldAddEndPuncttrue
\mciteSetBstMidEndSepPunct{\mcitedefaultmidpunct}
{\mcitedefaultendpunct}{\mcitedefaultseppunct}\relax
\EndOfBibitem
\bibitem{LHCB-PAPER-2014-041}
LHCb collaboration, R.~Aaij {\em et~al.},
  \ifthenelse{\boolean{articletitles}}{\emph{{Measurement of the CKM angle
  $\gamma$ using $\Bpm\to\D\Kpm$ with $\D\to\KS\pip\pim$, $\KS\Kp\Km$ decays}},
  }{}\href{http://dx.doi.org/10.1007/JHEP10(2014)097}{JHEP \textbf{10} (2014)
  097}, \href{http://arxiv.org/abs/1408.2748}{{\normalfont\ttfamily
  arXiv:1408.2748}}\relax
\mciteBstWouldAddEndPuncttrue
\mciteSetBstMidEndSepPunct{\mcitedefaultmidpunct}
{\mcitedefaultendpunct}{\mcitedefaultseppunct}\relax
\EndOfBibitem
\bibitem{LHCB-PAPER-2015-014}
LHCb collaboration, R.~Aaij {\em et~al.},
  \ifthenelse{\boolean{articletitles}}{\emph{{A study of $\CP$ violation in
  $\Bmp\to\D h^\mp$ $(h=K,\pi)$ with the modes $\D\to\Kmp\pipm\piz$,
  $\D\to\pip\pim\piz$ and $\D\to\Kp\Km\piz$}},
  }{}\href{http://dx.doi.org/10.1103/PhysRevD.91.112014}{Phys.\ Rev.\
  \textbf{D91} (2015) 112014},
  \href{http://arxiv.org/abs/1504.05442}{{\normalfont\ttfamily
  arXiv:1504.05442}}\relax
\mciteBstWouldAddEndPuncttrue
\mciteSetBstMidEndSepPunct{\mcitedefaultmidpunct}
{\mcitedefaultendpunct}{\mcitedefaultseppunct}\relax
\EndOfBibitem
\bibitem{GL}
M.~Gronau and D.~London, \ifthenelse{\boolean{articletitles}}{\emph{{How to
  determine all the angles of the unitarity triangle from $B^{0}_d \rightarrow
  D K_{S}$ and $B^{0}_{s}\rightarrow D \phi$}},
  }{}\href{http://dx.doi.org/http://dx.doi.org/10.1016/0370-2693(91)91756-L}{Phys.\
  Lett.\  \textbf{B253} (1991) 483}\relax
\mciteBstWouldAddEndPuncttrue
\mciteSetBstMidEndSepPunct{\mcitedefaultmidpunct}
{\mcitedefaultendpunct}{\mcitedefaultseppunct}\relax
\EndOfBibitem
\bibitem{GW}
M.~Gronau and D.~Wyler, \ifthenelse{\boolean{articletitles}}{\emph{{On
  determining a weak phase from charged B decay asymmetries }},
  }{}\href{http://dx.doi.org/http://dx.doi.org/10.1016/0370-2693(91)90034-N}{Phys.\
  Lett.\  \textbf{B265} (1991) 172}\relax
\mciteBstWouldAddEndPuncttrue
\mciteSetBstMidEndSepPunct{\mcitedefaultmidpunct}
{\mcitedefaultendpunct}{\mcitedefaultseppunct}\relax
\EndOfBibitem
\bibitem{ADS}
D.~Atwood, I.~Dunietz, and A.~Soni,
  \ifthenelse{\boolean{articletitles}}{\emph{{Enhanced \CP violation with $B
  \rightarrow KD^{0} ( \Dzb )$ modes and extraction of the
  Cabibbo-Kobayashi-Maskawa angle $\gamma$}},
  }{}\href{http://dx.doi.org/10.1103/PhysRevLett.78.3257}{Phys.\ Rev.\ Lett.\
  \textbf{78} (1997) 3257}\relax
\mciteBstWouldAddEndPuncttrue
\mciteSetBstMidEndSepPunct{\mcitedefaultmidpunct}
{\mcitedefaultendpunct}{\mcitedefaultseppunct}\relax
\EndOfBibitem
\bibitem{ADS-2001}
D.~Atwood, I.~Dunietz, and A.~Soni,
  \ifthenelse{\boolean{articletitles}}{\emph{{Improved methods for observing
  \CP violation in ${B}^\pm\rightarrow KD$ and measuring the CKM phase
  $\gamma$}}, }{}\href{http://dx.doi.org/10.1103/PhysRevD.63.036005}{Phys.\
  Rev.\  \textbf{D63} (2001) 036005},
  \href{http://arxiv.org/abs/hep-ph/0008090}{{\normalfont\ttfamily
  arXiv:hep-ph/0008090}}\relax
\mciteBstWouldAddEndPuncttrue
\mciteSetBstMidEndSepPunct{\mcitedefaultmidpunct}
{\mcitedefaultendpunct}{\mcitedefaultseppunct}\relax
\EndOfBibitem
\bibitem{LHCB-PAPER-2014-028}
LHCb collaboration, R.~Aaij {\em et~al.},
  \ifthenelse{\boolean{articletitles}}{\emph{{Measurement of $\CP$ violation
  parameters in $\Bz\to\D\Kstarz$ decays}},
  }{}\href{http://dx.doi.org/10.1103/PhysRevD.90.112002}{Phys.\ Rev.\
  \textbf{D90} (2014) 112002},
  \href{http://arxiv.org/abs/1407.8136}{{\normalfont\ttfamily
  arXiv:1407.8136}}\relax
\mciteBstWouldAddEndPuncttrue
\mciteSetBstMidEndSepPunct{\mcitedefaultmidpunct}
{\mcitedefaultendpunct}{\mcitedefaultseppunct}\relax
\EndOfBibitem
\bibitem{charminfo}
D.~Atwood and A.~Soni, \ifthenelse{\boolean{articletitles}}{\emph{Role of a
  charm factory in extracting {CKM}-phase information via ${B} \rightarrow
  \mathrm{DK}$}, }{}\href{http://dx.doi.org/10.1103/PhysRevD.68.033003}{Phys.\
  Rev.\  \textbf{D68} (2003) 033003},
  \href{http://arxiv.org/abs/hep-ph/0304085}{{\normalfont\ttfamily
  arXiv:hep-ph/0304085}}\relax
\mciteBstWouldAddEndPuncttrue
\mciteSetBstMidEndSepPunct{\mcitedefaultmidpunct}
{\mcitedefaultendpunct}{\mcitedefaultseppunct}\relax
\EndOfBibitem
\bibitem{charm4pi}
S.~Malde {\em et~al.}, \ifthenelse{\boolean{articletitles}}{\emph{First
  determination of the {CP} content of ${D} \rightarrow \pi^+\pi^-\pi^+\pi^-$
  and updated determination of the {CP} contents of ${D} \rightarrow
  \pi^+\pi^-\pi^0$ and ${D} \rightarrow {K}^+{K}^-\pi^0$},
  }{}\href{http://dx.doi.org/10.1016/j.physletb.2015.05.043}{Phys.\ Lett.\
  \textbf{B747} (2015) 9},
  \href{http://arxiv.org/abs/1504.05878}{{\normalfont\ttfamily
  arXiv:1504.05878}}\relax
\mciteBstWouldAddEndPuncttrue
\mciteSetBstMidEndSepPunct{\mcitedefaultmidpunct}
{\mcitedefaultendpunct}{\mcitedefaultseppunct}\relax
\EndOfBibitem
\bibitem{BaBarDKstar}
\babar collaboration, B.~Aubert {\em et~al.},
  \ifthenelse{\boolean{articletitles}}{\emph{{Measurement of \CP violation
  observables and parameters for the decays \decay{\Bpm}{\D\Kstarpm}}},
  }{}\href{http://dx.doi.org/10.1103/PhysRevD.80.092001}{Phys.\ Rev.\
  \textbf{D80} (2009) 092001},
  \href{http://arxiv.org/abs/0909.3981}{{\normalfont\ttfamily
  arXiv:0909.3981}}\relax
\mciteBstWouldAddEndPuncttrue
\mciteSetBstMidEndSepPunct{\mcitedefaultmidpunct}
{\mcitedefaultendpunct}{\mcitedefaultseppunct}\relax
\EndOfBibitem
\bibitem{BaBarGGSZ}
\babar collaboration, B.~Aubert {\em et~al.},
  \ifthenelse{\boolean{articletitles}}{\emph{{Improved measurement of the CKM
  angle $\ensuremath{\gamma}$ in
  ${B}^{\ensuremath{\mp}}\ensuremath{\rightarrow}{D}^{(*)}{K}^{(*)\ensuremath{\mp}}$
  decays with a Dalitz plot analysis of $D$ decays to
  ${K}_{S}^{0}{\ensuremath{\pi}}^{+}{\ensuremath{\pi}}^{-}$ and
  ${K}_{S}^{0}{K}^{+}{K}^{-}$}},
  }{}\href{http://dx.doi.org/10.1103/PhysRevD.78.034023}{Phys.\ Rev.\
  \textbf{D78} (2008) 034023},
  \href{http://arxiv.org/abs/0804.2089}{{\normalfont\ttfamily
  arXiv:0804.2089}}\relax
\mciteBstWouldAddEndPuncttrue
\mciteSetBstMidEndSepPunct{\mcitedefaultmidpunct}
{\mcitedefaultendpunct}{\mcitedefaultseppunct}\relax
\EndOfBibitem
\bibitem{BelleGGSZ}
Belle collaboration, A.~Poluektov {\em et~al.},
  \ifthenelse{\boolean{articletitles}}{\emph{{Measurement of
  ${\ensuremath{\phi}}_{3}$ with a Dalitz plot analysis of
  ${B}^{+}\ensuremath{\rightarrow}{D}^{(*)}{K}^{(*)+}$ decay}},
  }{}\href{http://dx.doi.org/10.1103/PhysRevD.73.112009}{Phys.\ Rev.\
  \textbf{D73} (2006) 112009},
  \href{http://arxiv.org/abs/hep-ex/0604054}{{\normalfont\ttfamily
  arXiv:hep-ex/0604054}}\relax
\mciteBstWouldAddEndPuncttrue
\mciteSetBstMidEndSepPunct{\mcitedefaultmidpunct}
{\mcitedefaultendpunct}{\mcitedefaultseppunct}\relax
\EndOfBibitem
\bibitem{charmcpv}
W.~Wang, \ifthenelse{\boolean{articletitles}}{\emph{{\CP violation effects on
  the measurement of the Cabibbo-Kobayashi-Maskawa angle $\gamma$ from ${B}
  \rightarrow {DK}$}},
  }{}\href{http://dx.doi.org/10.1103/PhysRevLett.110.061802}{Phys.\ Rev.\
  Lett.\  \textbf{110} (2013) 061802},
  \href{http://arxiv.org/abs/1211.4539}{{\normalfont\ttfamily
  arXiv:1211.4539}}\relax
\mciteBstWouldAddEndPuncttrue
\mciteSetBstMidEndSepPunct{\mcitedefaultmidpunct}
{\mcitedefaultendpunct}{\mcitedefaultseppunct}\relax
\EndOfBibitem
\bibitem{Gronau2003198}
M.~Gronau, \ifthenelse{\boolean{articletitles}}{\emph{{Improving bounds on
  $\gamma$ in $B^{\pm} \to DK^{*\pm}$ and $B^{\pm,0} \to D X_s^{\pm,0}$}},
  }{}\href{http://dx.doi.org/http://dx.doi.org/10.1016/S0370-2693(03)00192-8}{Phys.\
  Lett.\  \textbf{B557} (2003) 198}\relax
\mciteBstWouldAddEndPuncttrue
\mciteSetBstMidEndSepPunct{\mcitedefaultmidpunct}
{\mcitedefaultendpunct}{\mcitedefaultseppunct}\relax
\EndOfBibitem
\bibitem{charmmixing}
M.~Rama, \ifthenelse{\boolean{articletitles}}{\emph{Effect of
  ${D}\ensuremath{-}\bar{D}$ mixing in the extraction of $\ensuremath{\gamma}$
  with
  ${B}^{\ensuremath{-}}\ensuremath{\rightarrow}{D}^{0}{K}^{\ensuremath{-}}$ and
  ${B}^{\ensuremath{-}}\ensuremath{\rightarrow}{D}^{0}{\ensuremath{\pi}}^{\ensuremath{-}}$
  decays}, }{}\href{http://dx.doi.org/10.1103/PhysRevD.89.014021}{Phys.\ Rev.\
  \textbf{D89} (2014) 014021},
  \href{http://arxiv.org/abs/1307.4384}{{\normalfont\ttfamily
  arXiv:1307.4384}}\relax
\mciteBstWouldAddEndPuncttrue
\mciteSetBstMidEndSepPunct{\mcitedefaultmidpunct}
{\mcitedefaultendpunct}{\mcitedefaultseppunct}\relax
\EndOfBibitem
\bibitem{HFLAV16}
Heavy Flavor Averaging Group, Y.~Amhis {\em et~al.},
  \ifthenelse{\boolean{articletitles}}{\emph{{Averages of $b$-hadron,
  $c$-hadron, and $\tau$-lepton properties as of summer 2016}},
  }{}\href{http://arxiv.org/abs/1612.07233}{{\normalfont\ttfamily
  arXiv:1612.07233}}, {updated results and plots available at
  \href{http://www.slac.stanford.edu/xorg/hflav/}{{\texttt{http://www.slac.stanford.edu/xorg/hflav/}}}}\relax
\mciteBstWouldAddEndPuncttrue
\mciteSetBstMidEndSepPunct{\mcitedefaultmidpunct}
{\mcitedefaultendpunct}{\mcitedefaultseppunct}\relax
\EndOfBibitem
\bibitem{charmk3pi}
T.~Evans {\em et~al.}, \ifthenelse{\boolean{articletitles}}{\emph{{Improved
  determination of the ${D} \rightarrow {K}^+\pi^-\pi^+\pi^-$ coherence factor
  and associated hadronic parameters from a combination of $e^+e^- \rightarrow
  \Upsilon(3770) \rightarrow c\bar{c}$ and $pp \rightarrow c\bar{c}{X}$ data}},
  }{}\href{http://dx.doi.org/10.1016/j.physletb.2016.04.037}{Phys.\ Lett.\
  \textbf{B757} (2016) 520}, Corrigendum
  \href{http://dx.doi.org/10.1106/j.physletb.2016.11.021}{ibid.\
  \textbf{B765} (2017) 402},
  \href{http://arxiv.org/abs/1602.07430}{{\normalfont\ttfamily
  arXiv:1602.07430}}\relax
\mciteBstWouldAddEndPuncttrue
\mciteSetBstMidEndSepPunct{\mcitedefaultmidpunct}
{\mcitedefaultendpunct}{\mcitedefaultseppunct}\relax
\EndOfBibitem
\bibitem{LHCB-PAPER-2015-057}
LHCb collaboration, R.~Aaij {\em et~al.},
  \ifthenelse{\boolean{articletitles}}{\emph{{First observation of $\Dz-\Dzb$
  oscillations in $\Dz\to\Kp\pip\pim\pim$ decays and a measurement of the
  associated coherence parameters}},
  }{}\href{http://dx.doi.org/10.1103/PhysRevLett.116.241801}{Phys.\ Rev.\
  Lett.\  \textbf{116} (2016) 241801},
  \href{http://arxiv.org/abs/1602.07224}{{\normalfont\ttfamily
  arXiv:1602.07224}}\relax
\mciteBstWouldAddEndPuncttrue
\mciteSetBstMidEndSepPunct{\mcitedefaultmidpunct}
{\mcitedefaultendpunct}{\mcitedefaultseppunct}\relax
\EndOfBibitem
\bibitem{Alves:2008zz}
LHCb collaboration, A.~A. Alves~Jr.\ {\em et~al.},
  \ifthenelse{\boolean{articletitles}}{\emph{{The \lhcb detector at the LHC}},
  }{}\href{http://dx.doi.org/10.1088/1748-0221/3/08/S08005}{JINST \textbf{3}
  (2008) S08005}\relax
\mciteBstWouldAddEndPuncttrue
\mciteSetBstMidEndSepPunct{\mcitedefaultmidpunct}
{\mcitedefaultendpunct}{\mcitedefaultseppunct}\relax
\EndOfBibitem
\bibitem{LHCb-DP-2014-002}
LHCb collaboration, R.~Aaij {\em et~al.},
  \ifthenelse{\boolean{articletitles}}{\emph{{LHCb detector performance}},
  }{}\href{http://dx.doi.org/10.1142/S0217751X15300227}{Int.\ J.\ Mod.\ Phys.\
  \textbf{A30} (2015) 1530022},
  \href{http://arxiv.org/abs/1412.6352}{{\normalfont\ttfamily
  arXiv:1412.6352}}\relax
\mciteBstWouldAddEndPuncttrue
\mciteSetBstMidEndSepPunct{\mcitedefaultmidpunct}
{\mcitedefaultendpunct}{\mcitedefaultseppunct}\relax
\EndOfBibitem
\bibitem{LHCb-DP-2012-004}
R.~Aaij {\em et~al.}, \ifthenelse{\boolean{articletitles}}{\emph{{The \lhcb
  trigger and its performance in 2011}},
  }{}\href{http://dx.doi.org/10.1088/1748-0221/8/04/P04022}{JINST \textbf{8}
  (2013) P04022}, \href{http://arxiv.org/abs/1211.3055}{{\normalfont\ttfamily
  arXiv:1211.3055}}\relax
\mciteBstWouldAddEndPuncttrue
\mciteSetBstMidEndSepPunct{\mcitedefaultmidpunct}
{\mcitedefaultendpunct}{\mcitedefaultseppunct}\relax
\EndOfBibitem
\bibitem{BBDT}
V.~V. Gligorov and M.~Williams,
  \ifthenelse{\boolean{articletitles}}{\emph{{Efficient, reliable and fast
  high-level triggering using a bonsai boosted decision tree}},
  }{}\href{http://dx.doi.org/10.1088/1748-0221/8/02/P02013}{JINST \textbf{8}
  (2013) P02013}, \href{http://arxiv.org/abs/1210.6861}{{\normalfont\ttfamily
  arXiv:1210.6861}}\relax
\mciteBstWouldAddEndPuncttrue
\mciteSetBstMidEndSepPunct{\mcitedefaultmidpunct}
{\mcitedefaultendpunct}{\mcitedefaultseppunct}\relax
\EndOfBibitem
\bibitem{LHCB-PAPER-2015-037}
LHCb collaboration, R.~Aaij {\em et~al.},
  \ifthenelse{\boolean{articletitles}}{\emph{{Measurement of forward $\jpsi$
  production cross-sections in $\proton\proton$ collisions at
  $\sqrt{s}=13$\tev}}, }{}\href{http://dx.doi.org/10.1007/JHEP10(2015)172}{JHEP
  \textbf{10} (2015) 172}, Erratum
  \href{http://dx.doi.org/10.1007/JHEP05(2017)063}{ibid.\   \textbf{05} (2017)
  063}, \href{http://arxiv.org/abs/1509.00771}{{\normalfont\ttfamily
  arXiv:1509.00771}}\relax
\mciteBstWouldAddEndPuncttrue
\mciteSetBstMidEndSepPunct{\mcitedefaultmidpunct}
{\mcitedefaultendpunct}{\mcitedefaultseppunct}\relax
\EndOfBibitem
\bibitem{LHCb-DP-2012-003}
M.~Adinolfi {\em et~al.},
  \ifthenelse{\boolean{articletitles}}{\emph{{Performance of the \lhcb RICH
  detector at the LHC}},
  }{}\href{http://dx.doi.org/10.1140/epjc/s10052-013-2431-9}{Eur.\ Phys.\ J.\
  \textbf{C73} (2013) 2431},
  \href{http://arxiv.org/abs/1211.6759}{{\normalfont\ttfamily
  arXiv:1211.6759}}\relax
\mciteBstWouldAddEndPuncttrue
\mciteSetBstMidEndSepPunct{\mcitedefaultmidpunct}
{\mcitedefaultendpunct}{\mcitedefaultseppunct}\relax
\EndOfBibitem
\bibitem{Sjostrand:2006za}
T.~Sj\"{o}strand, S.~Mrenna, and P.~Skands,
  \ifthenelse{\boolean{articletitles}}{\emph{{PYTHIA 6.4 physics and manual}},
  }{}\href{http://dx.doi.org/10.1088/1126-6708/2006/05/026}{JHEP \textbf{05}
  (2006) 026}, \href{http://arxiv.org/abs/hep-ph/0603175}{{\normalfont\ttfamily
  arXiv:hep-ph/0603175}}\relax
\mciteBstWouldAddEndPuncttrue
\mciteSetBstMidEndSepPunct{\mcitedefaultmidpunct}
{\mcitedefaultendpunct}{\mcitedefaultseppunct}\relax
\EndOfBibitem
\bibitem{Sjostrand:2007gs}
T.~Sj\"{o}strand, S.~Mrenna, and P.~Skands,
  \ifthenelse{\boolean{articletitles}}{\emph{{A brief introduction to PYTHIA
  8.1}}, }{}\href{http://dx.doi.org/10.1016/j.cpc.2008.01.036}{Comput.\ Phys.\
  Commun.\  \textbf{178} (2008) 852},
  \href{http://arxiv.org/abs/0710.3820}{{\normalfont\ttfamily
  arXiv:0710.3820}}\relax
\mciteBstWouldAddEndPuncttrue
\mciteSetBstMidEndSepPunct{\mcitedefaultmidpunct}
{\mcitedefaultendpunct}{\mcitedefaultseppunct}\relax
\EndOfBibitem
\bibitem{LHCb-PROC-2010-056}
I.~Belyaev {\em et~al.}, \ifthenelse{\boolean{articletitles}}{\emph{{Handling
  of the generation of primary events in Gauss, the LHCb simulation
  framework}}, }{}\href{http://dx.doi.org/10.1088/1742-6596/331/3/032047}{{J.\
  Phys.\ Conf.\ Ser.\ } \textbf{331} (2011) 032047}\relax
\mciteBstWouldAddEndPuncttrue
\mciteSetBstMidEndSepPunct{\mcitedefaultmidpunct}
{\mcitedefaultendpunct}{\mcitedefaultseppunct}\relax
\EndOfBibitem
\bibitem{Lange:2001uf}
D.~J. Lange, \ifthenelse{\boolean{articletitles}}{\emph{{The EvtGen particle
  decay simulation package}},
  }{}\href{http://dx.doi.org/10.1016/S0168-9002(01)00089-4}{Nucl.\ Instrum.\
  Meth.\  \textbf{A462} (2001) 152}\relax
\mciteBstWouldAddEndPuncttrue
\mciteSetBstMidEndSepPunct{\mcitedefaultmidpunct}
{\mcitedefaultendpunct}{\mcitedefaultseppunct}\relax
\EndOfBibitem
\bibitem{Golonka:2005pn}
P.~Golonka and Z.~Was, \ifthenelse{\boolean{articletitles}}{\emph{{PHOTOS Monte
  Carlo: A precision tool for QED corrections in $Z$ and $W$ decays}},
  }{}\href{http://dx.doi.org/10.1140/epjc/s2005-02396-4}{Eur.\ Phys.\ J.\
  \textbf{C45} (2006) 97},
  \href{http://arxiv.org/abs/hep-ph/0506026}{{\normalfont\ttfamily
  arXiv:hep-ph/0506026}}\relax
\mciteBstWouldAddEndPuncttrue
\mciteSetBstMidEndSepPunct{\mcitedefaultmidpunct}
{\mcitedefaultendpunct}{\mcitedefaultseppunct}\relax
\EndOfBibitem
\bibitem{Allison:2006ve}
Geant4 collaboration, J.~Allison {\em et~al.},
  \ifthenelse{\boolean{articletitles}}{\emph{{Geant4 developments and
  applications}}, }{}\href{http://dx.doi.org/10.1109/TNS.2006.869826}{IEEE
  Trans.\ Nucl.\ Sci.\  \textbf{53} (2006) 270}\relax
\mciteBstWouldAddEndPuncttrue
\mciteSetBstMidEndSepPunct{\mcitedefaultmidpunct}
{\mcitedefaultendpunct}{\mcitedefaultseppunct}\relax
\EndOfBibitem
\bibitem{Agostinelli:2002hh}
Geant4 collaboration, S.~Agostinelli {\em et~al.},
  \ifthenelse{\boolean{articletitles}}{\emph{{Geant4: A simulation toolkit}},
  }{}\href{http://dx.doi.org/10.1016/S0168-9002(03)01368-8}{Nucl.\ Instrum.\
  Meth.\  \textbf{A506} (2003) 250}\relax
\mciteBstWouldAddEndPuncttrue
\mciteSetBstMidEndSepPunct{\mcitedefaultmidpunct}
{\mcitedefaultendpunct}{\mcitedefaultseppunct}\relax
\EndOfBibitem
\bibitem{LHCb-PROC-2011-006}
M.~Clemencic {\em et~al.}, \ifthenelse{\boolean{articletitles}}{\emph{{The
  \lhcb simulation application, Gauss: Design, evolution and experience}},
  }{}\href{http://dx.doi.org/10.1088/1742-6596/331/3/032023}{{J.\ Phys.\ Conf.\
  Ser.\ } \textbf{331} (2011) 032023}\relax
\mciteBstWouldAddEndPuncttrue
\mciteSetBstMidEndSepPunct{\mcitedefaultmidpunct}
{\mcitedefaultendpunct}{\mcitedefaultseppunct}\relax
\EndOfBibitem
\bibitem{PDG2016}
Particle Data Group, C.~Patrignani {\em et~al.},
  \ifthenelse{\boolean{articletitles}}{\emph{{\href{http://pdg.lbl.gov/}{Review
  of particle physics}}},
  }{}\href{http://dx.doi.org/10.1088/1674-1137/40/10/100001}{Chin.\ Phys.\
  \textbf{C40} (2016) 100001}\relax
\mciteBstWouldAddEndPuncttrue
\mciteSetBstMidEndSepPunct{\mcitedefaultmidpunct}
{\mcitedefaultendpunct}{\mcitedefaultseppunct}\relax
\EndOfBibitem
\bibitem{Hulsbergen:2005pu}
W.~D. Hulsbergen, \ifthenelse{\boolean{articletitles}}{\emph{{Decay chain
  fitting with a Kalman filter}},
  }{}\href{http://dx.doi.org/10.1016/j.nima.2005.06.078}{Nucl.\ Instrum.\
  Meth.\  \textbf{A552} (2005) 566},
  \href{http://arxiv.org/abs/physics/0503191}{{\normalfont\ttfamily
  arXiv:physics/0503191}}\relax
\mciteBstWouldAddEndPuncttrue
\mciteSetBstMidEndSepPunct{\mcitedefaultmidpunct}
{\mcitedefaultendpunct}{\mcitedefaultseppunct}\relax
\EndOfBibitem
\bibitem{Breiman}
L.~Breiman, J.~H. Friedman, R.~A. Olshen, and C.~J. Stone, {\em Classification
  and regression trees}, Wadsworth international group, Belmont, California,
  USA, 1984\relax
\mciteBstWouldAddEndPuncttrue
\mciteSetBstMidEndSepPunct{\mcitedefaultmidpunct}
{\mcitedefaultendpunct}{\mcitedefaultseppunct}\relax
\EndOfBibitem
\bibitem{Skwarnicki:1986xj}
T.~Skwarnicki, {\em {A study of the radiative cascade transitions between the
  Upsilon-prime and Upsilon resonances}}, PhD thesis, Institute of Nuclear
  Physics, Krakow, 1986,
  {\href{http://inspirehep.net/record/230779/}{DESY-F31-86-02}}\relax
\mciteBstWouldAddEndPuncttrue
\mciteSetBstMidEndSepPunct{\mcitedefaultmidpunct}
{\mcitedefaultendpunct}{\mcitedefaultseppunct}\relax
\EndOfBibitem
\bibitem{LHCB-PAPER-2017-021}
LHCb collaboration, R.~Aaij {\em et~al.},
  \ifthenelse{\boolean{articletitles}}{\emph{{Measurement of $C\!P$ observables
  in $B^\pm \to D^{(\ast)}K^\pm$ and $B^\pm \to D^{(\ast)}\pi^\pm$ decays}},
  }{}\href{http://arxiv.org/abs/1708.06370}{{\normalfont\ttfamily
  arXiv:1708.06370}}, {submitted to Phys. Lett. B}\relax
\mciteBstWouldAddEndPuncttrue
\mciteSetBstMidEndSepPunct{\mcitedefaultmidpunct}
{\mcitedefaultendpunct}{\mcitedefaultseppunct}\relax
\EndOfBibitem
\bibitem{LHCB-PAPER-2016-006}
LHCb collaboration, R.~Aaij {\em et~al.},
  \ifthenelse{\boolean{articletitles}}{\emph{{Model-independent measurement of
  the CKM angle $\gamma$ using $\Bz \to \D\Kstarz$ decays with
  $\D\to\KS\pip\pim$ and $\KS\Kp\Km$}},
  }{}\href{http://dx.doi.org/10.1007/JHEP06(2016)131}{JHEP \textbf{06} (2016)
  131}, \href{http://arxiv.org/abs/1604.01525}{{\normalfont\ttfamily
  arXiv:1604.01525}}\relax
\mciteBstWouldAddEndPuncttrue
\mciteSetBstMidEndSepPunct{\mcitedefaultmidpunct}
{\mcitedefaultendpunct}{\mcitedefaultseppunct}\relax
\EndOfBibitem
\bibitem{LHCb-PAPER-2016-054}
LHCb collaboration, R.~Aaij {\em et~al.},
  \ifthenelse{\boolean{articletitles}}{\emph{{Measurement of the $B^\pm$
  production asymmetry and the $C\!P$ asymmetry in $B^\pm\to \jpsi K^\pm$
  decays}}, }{}\href{http://dx.doi.org/10.1103/PhysRevD.95.052005}{Phys.\ Rev.\
   \textbf{D95} (2017) 052005},
  \href{http://arxiv.org/abs/1701.05501}{{\normalfont\ttfamily
  arXiv:1701.05501}}\relax
\mciteBstWouldAddEndPuncttrue
\mciteSetBstMidEndSepPunct{\mcitedefaultmidpunct}
{\mcitedefaultendpunct}{\mcitedefaultseppunct}\relax
\EndOfBibitem
\bibitem{LHCb-PAPER-2014-013}
LHCb collaboration, R.~Aaij {\em et~al.},
  \ifthenelse{\boolean{articletitles}}{\emph{{Measurement of $\CP$ asymmetry in
  $\Dz\to\Km\Kp$ and $\Dz\to\pim\pip$ decays}},
  }{}\href{http://dx.doi.org/10.1007/JHEP07(2014)041}{JHEP \textbf{07} (2014)
  041}, \href{http://arxiv.org/abs/1405.2797}{{\normalfont\ttfamily
  arXiv:1405.2797}}\relax
\mciteBstWouldAddEndPuncttrue
\mciteSetBstMidEndSepPunct{\mcitedefaultmidpunct}
{\mcitedefaultendpunct}{\mcitedefaultseppunct}\relax
\EndOfBibitem
\bibitem{LHCB-PAPER-2012-009}
LHCb collaboration, R.~Aaij {\em et~al.},
  \ifthenelse{\boolean{articletitles}}{\emph{{Measurement of the $\Dsp$--$\Dsm$
  production asymmetry in $7$\tev $\proton\proton$ collisions}},
  }{}\href{http://dx.doi.org/10.1016/j.physletb.2012.06.001}{Phys.\ Lett.\
  \textbf{B713} (2012) 186},
  \href{http://arxiv.org/abs/1205.0897}{{\normalfont\ttfamily
  arXiv:1205.0897}}\relax
\mciteBstWouldAddEndPuncttrue
\mciteSetBstMidEndSepPunct{\mcitedefaultmidpunct}
{\mcitedefaultendpunct}{\mcitedefaultseppunct}\relax
\EndOfBibitem
\bibitem{Wilks:1938dza}
S.~S. Wilks, \ifthenelse{\boolean{articletitles}}{\emph{{The large-sample
  distribution of the likelihood ratio for testing composite hypotheses}},
  }{}\href{http://dx.doi.org/10.1214/aoms/1177732360}{Ann.\ Math.\ Stat.\
  \textbf{9} (1938) 60}\relax
\mciteBstWouldAddEndPuncttrue
\mciteSetBstMidEndSepPunct{\mcitedefaultmidpunct}
{\mcitedefaultendpunct}{\mcitedefaultseppunct}\relax
\EndOfBibitem
\bibitem{LHCB-PAPER-2014-036}
LHCb collaboration, R.~Aaij {\em et~al.},
  \ifthenelse{\boolean{articletitles}}{\emph{{Dalitz plot analysis of
  $\Bs\to\Dzb\Km\pip$ decays}},
  }{}\href{http://dx.doi.org/10.1103/PhysRevD.90.072003}{Phys.\ Rev.\
  \textbf{D90} (2014) 072003},
  \href{http://arxiv.org/abs/1407.7712}{{\normalfont\ttfamily
  arXiv:1407.7712}}\relax
\mciteBstWouldAddEndPuncttrue
\mciteSetBstMidEndSepPunct{\mcitedefaultmidpunct}
{\mcitedefaultendpunct}{\mcitedefaultseppunct}\relax
\EndOfBibitem
\bibitem{Laura}
T.~Latham, J.~Back, and P.~Harrison,
  \ifthenelse{\boolean{articletitles}}{\emph{{Laura++, a Dalitz plot fitting
  package}}, }{}
\newblock {available at \url{https://laura.hepforge.org/}}\relax
\mciteBstWouldAddEndPuncttrue
\mciteSetBstMidEndSepPunct{\mcitedefaultmidpunct}
{\mcitedefaultendpunct}{\mcitedefaultseppunct}\relax
\EndOfBibitem
\bibitem{LASS}
D.~Aston {\em et~al.}, \ifthenelse{\boolean{articletitles}}{\emph{{A study of
  $K^-\pi^+$ scattering in the reaction $K^-p \to K^-\pi^+n$ at 11 GeV/c}},
  }{}\href{http://dx.doi.org/http://dx.doi.org/10.1016/0550-3213(88)90028-4}{Nucl.\
  Phys.\  \textbf{B296} (1988) 493}\relax
\mciteBstWouldAddEndPuncttrue
\mciteSetBstMidEndSepPunct{\mcitedefaultmidpunct}
{\mcitedefaultendpunct}{\mcitedefaultseppunct}\relax
\EndOfBibitem
\end{mcitethebibliography}

\newpage

\centerline{\large\bf LHCb collaboration}
\begin{flushleft}
\small
R.~Aaij$^{40}$,
B.~Adeva$^{39}$,
M.~Adinolfi$^{48}$,
Z.~Ajaltouni$^{5}$,
S.~Akar$^{59}$,
J.~Albrecht$^{10}$,
F.~Alessio$^{40}$,
M.~Alexander$^{53}$,
A.~Alfonso~Albero$^{38}$,
S.~Ali$^{43}$
% ...

{\footnotesize \it
% ...
$ ^{5}$Clermont Universit{\'e}, Universit{\'e} Blaise Pascal, CNRS/IN2P3, LPC, Clermont-Ferrand, France\\
% ...
$ ^{10}$Fakult{\"a}t Physik, Technische Universit{\"a}t Dortmund, Dortmund, Germany\\
% ...
$ ^{38}$ICCUB, Universitat de Barcelona, Barcelona, Spain\\
$ ^{39}$Universidad de Santiago de Compostela, Santiago de Compostela, Spain\\
$ ^{40}$European Organization for Nuclear Research (CERN), Geneva, Switzerland\\
% ...
$ ^{43}$Nikhef National Institute for Subatomic Physics, Amsterdam, The Netherlands\\
% ...
$ ^{48}$H.H. Wills Physics Laboratory, University of Bristol, Bristol, United Kingdom\\
% ...
$ ^{53}$School of Physics and Astronomy, University of Glasgow, Glasgow, United Kingdom\\
% ...
$ ^{59}$University of Cincinnati, Cincinnati, OH, United States\\
% ...
}
\end{flushleft}

\end{document}